\documentclass[12pt]{article}

\usepackage{amsmath}
\usepackage{graphicx}
\usepackage{booktabs}
\usepackage{amsfonts}
\usepackage{hyperref}
\usepackage{amssymb}
\usepackage{caption}
\usepackage{subcaption}
\usepackage{color}

\renewcommand{\theequation}
{\arabic{section}.\arabic{equation}}

\makeatletter
\def\eqnarray{ \stepcounter{equation} \let\@currentlabel=\theequation
 \global\@eqnswtrue
 \global\@eqcnt\z@
 \tabskip\@centering
 \let\\=\@eqncr
 $$\halign to \displaywidth\bgroup\@eqnsel\hskip\@centering
 $\displaystyle\tabskip\z@{##}$&\global\@eqcnt\@ne
 \hfil$\displaystyle{{}##{}}$\hfil
 &\global\@eqcnt\tw@$\displaystyle\tabskip\z@{##}$\hfil
 \tabskip\@centering&\llap{##}\tabskip\z@\cr}
\makeatother

\makeatletter
\def\@arrayacol{\edef\@preamble{\@preamble \hskip .5\arraycolsep}}
\def\array{\let\@acol\@arrayacol \let\@classz\@arrayclassz
\let\@classiv\@arrayclassiv \let\\\@arraycr\def\@halignto{}\@tabarray}
\makeatother

\makeatletter
\newcounter{subeqncnt}
\def\thesubeqncnt{\alph{subeqncnt}}
\def\subequations{\begingroup%
   \stepcounter{equation}\edef\@tempa{\theequation}%
   \let\c@equation\c@subeqncnt\c@subeqncnt\z@
   \edef\theequation{\@tempa\noexpand\thesubeqncnt}}

\makeatother

\newcommand{\be}{\begin{equation}}
\newcommand{\ee}{\end{equation}}

\newcommand{\bea}{\begin{eqnarray}}
\newcommand{\eea}{\end{eqnarray}}

\def\CD {{\cal D}}

\def\CL {{\cal L}}

\def\CN {{\cal N}}
\def\cN {{\cal N}}

\def\Det{{\rm Det}}

\newcommand{\tr}{{\rm Tr}\,}
\newcommand{\dd}{{\rm d}}

\begin{document}

\setlength{\baselineskip}{7mm}
\begin{titlepage}
 \begin{flushright}
{\tt NRCPS-HE-52-2016}
\end{flushright}

\begin{center}
{\Large ~\\{\it   Distribution of Periodic Trajectories of Anosov C-system
\vspace{1cm}

}

}

\vspace{1cm}

{\sl Andrzej G\"orlich${}^{b,c}$, Marios Kalomenopoulos${}^a$, Konstantin Savvidy${}^a$ and George Savvidy${}^a$

\bigskip
${}^{a}$ \sl Institute of Nuclear and Particle Physics\\ Demokritos National Research Center, Ag. Paraskevi,  Athens, Greece\\
${}^{b}$ \sl The Niels Bohr Institute, Copenhagen University, Blegdamsvej 17, DK-2100 Copenhagen \O, Denmark\\
{${}^{c}$ \sl Institute of Physics, Jagiellonian University, Lojasiewicza 11, PL 30-348 Krakow, Poland}\\
\bigskip

}
\end{center}
\vspace{30pt}

\centerline{{\bf Abstract}}
The hyperbolic Anosov C-systems have a countable set of everywhere dense periodic trajectories which have been recently used to generate pseudorandom numbers.  
The asymptotic distribution of periodic trajectories of C-systems with periods less than a 
given number is well known, but a deviation of this distribution from its asymptotic behaviour is less known. Using 
fast algorithms, we are studying the exact distribution of periodic 
trajectories  and their deviation from asymptotic behaviour 
for hyperbolic C-systems which are defined on high dimensional  tori  and are used for Monte-Carlo simulations. A particular C-system which we consider in this article is the one which was implemented in the MIXMAX generator of pseudorandom numbers. The generator has the best combination of speed, reasonable size of the state, and availability for implementing the parallelization and is currently available  generator in the ROOT and CLHEP software packages at CERN.

\noindent

\end{titlepage}




\pagestyle{plain}

\section{\it Introduction  }

We shall consider a dynamical system $T$ with the phase space $W^N$
which is an $N$-dimensional torus appearing at  factorisation of the
Euclidean space $E^N$ with coordinates $ (w_1,...,w_N)$ over an integer lattice.
The automorphisms of the torus are generated by the linear transformation
\bea\label{cmap}
w_i \rightarrow \sum T_{i,j} w_j,~~~~(mod ~1),
\eea
where the integer matrix $T$ has a determinant equal to one: $Det T =1$.
The automorphism  $T$ can be thought of as a linear transformation of the Euclidean space $E^N$
which preserves the lattice $\CL$ of points with integer coordinates.

In order for the automorphisms of the torus (\ref{cmap})  {\it to fulfil  the Anosov C-condition it is necessary
and sufficient that the matrix $T$ has no eigenvalues on the unit circle} \cite{anosov}.
Thus the  spectrum $\{ \Lambda = {\lambda_1},...,
\lambda_N \}$ of the matrix $T$ is chosen to fulfil the following
two conditions \cite{anosov}:
\bea\label{mmatrix}
1)~Det T=  {\lambda_1}{\lambda_2}....{\lambda_N}=1,~~~~~
2)~\vert {\lambda_i} \vert \neq 1, ~~~~~~~~~~~\forall i.
\eea
Because the determinant of the matrix $T$ is equal to one,
the Liouville's measure $d\mu = dx_1...dx_N$ is invariant under the action of $T$.
The inverse matrix $T^{-1}$ is also an integer matrix because $Det T=1$.
Therefore $T$ is an automorphism of  the torus $W^N$  onto itself.
This automorphism has a fixed point $w=0$ corresponding to the origin of $E^N$.The above conditions (\ref{mmatrix}) on the eigenvalues of the matrix $T$ are  sufficient
to prove that the system belongs to the class of  Anosov C-systems \cite{anosov}.  

The C-systems have a countable set of everywhere dense 
periodic trajectories \cite{anosov} and in  this paper we shall  study the distribution functions of the periodic trajectories 
of  C-system defined on a 
torus.   Earlier the properties of Yang-Mills 
mechanics and their relation with the C- and K- systems 
were studied in \cite{yangmillsmech,Savvidy:1982jk}, see also the recent 
articles \cite{Gur-Ari:2015rcq,Brown:2016wib}.
Later on in \cite{yer1986a,konstantin,Savvidy:2015ida,Savvidy:2015jva}
 the C-systems on tori have been suggested to generate pseudorandom numbers for Monte-Carlo simulations and it is therefore important to know the properties of the corresponding 
periodic trajectories. 

It is convenient to divide the eigenvalues of the matrix $T$
into two sets   $\{ \lambda_{\alpha}  \} $ and $\{  \lambda_{\beta }  \} $
with modulus smaller and larger than one:
\bea\label{eigenvalues}
0 <  \vert \lambda_{\alpha} \vert   < 1 <
\vert \lambda_{\beta}\vert.
\eea
These is useful in calculating the entropy of the C-systems
  on tori \cite{anosov,kolmo,kolmo1,sinai3,rokhlin1,rokhlin2,sinai4,gines}.
 The most convenient way to calculate the entropy of a C-system is to integrate over the whole
phase space the logarithm of the volume expansion rate $\lambda(w)$
of a $l$-dimensional infinitesimal cube which is embedded  into the expanding foliation \cite{anosov,sinai3,rokhlin1,rokhlin2,sinai4,gines}:
\be\label{biuty}
h(T) = \int_{W^N} \ln \lambda(w) d w.
\ee
Here the volume of the $W^N$ is normalised to 1. For the automorphisms
on a torus (\ref{cmap}) the coefficient $\lambda(w)$  does not depend on the phase
space coordinates $w$ and is equal to the product of eigenvalues
$\{  \lambda_{\beta }  \} $  with modulus  larger than one:
\be\label{entropyofT}
h(T) = \sum_{\vert \lambda_{\beta} \vert > 1} \ln \vert \lambda_{\beta} \vert.
\ee
It was discovered in \cite{anosov,margulis,bowen0,bowen,bowen1} that the entropy measures  
not only the ergodic and stochastic properties of the dynamical systems, 
but also the variety  and richness of the periodic trajectories of the C-systems \cite{anosov,smale,margulis,bowen0,bowen,bowen1}. 
The C-systems have a countable set of everywhere dense 
periodic trajectories \cite{anosov} and their density grows exponentially 
with entropy
\cite{anosov,margulis,bowen0,bowen,bowen1}.  Our aim is to 
study the behaviour of these densities in the next to leading order.

{\it All trajectories with rational coordinates $(w_1,...,w_N)$, and only they,
are periodic trajectories of the automorphisms of the  torus} (\ref{cmap}).
Let us fix the integer number $p$, then the points on a torus with
the coordinates having a denominator $p$ form a rational lattice  $\CL_p$ 
with coordinates $\{a_1/p,...,a_N/p \}$, where $a_i =0,1,...,p-1$. The
automorphism (\ref{mmatrix}) with integer entries transforms the
points of the rational lattice $\CL_p$  into themselves, therefore all these points belong to periodic trajectories. 

Conversely, 
if  the point $w=(w_1,...,w_N)$ belongs to a trajectory of period $n > 1$,  then
 \be\label{periodictrajectories1}
 T^n w = w + b,
 \ee
 where $b$ is an integer vector. The above equation with
 respect to $w$ has the  matrix $T^n -1$ with a nonzero determinant, 
 therefore the components of $w$ are {\it rational}.
Thus the periodic trajectories of the period $n$ of the automorphism $T$ are given
by the solution of the equation (\ref{periodictrajectories1}),
where $b \in Z^N$ is an integer vector.
 
 As $b$ varies in $Z^N$, the solutions of the equation (\ref{periodictrajectories1}) determine
 a fundamental domain $\CD_n$ in the covering Euclidian space $E^N$  of the volume
 $\mu(D_n)=1/\vert Det (T^n-1) \vert $. Therefore the number of all points $\CN_n$ on the
 periodic trajectories of the period $n$
 is given by  the corresponding inverse volume \cite{smale,margulis,bowen0,bowen,bowen1}:
 \be\label{numbers}
\CN_n = \vert Det (T^n-1) \vert = \vert  \prod^{N}_{i=1}(\lambda^n_i -1) \vert .
 \ee
The Bowen theorem \cite{bowen,bowen1} states that the entropy
of the automorphism $T$ can be equivalently represented  in terms of $\CN_n$:
\be\label{bowen}
h(T)= \lim_{n \rightarrow \infty}{1\over n}~ \ln  \CN_n=
\lim_{n \rightarrow \infty}{1\over n} \ln (\vert  \prod^{N}_{i=1}(\lambda^n_i -1) \vert) =
\sum_{\vert \lambda_{\beta} \vert > 1} \ln \vert \lambda_{\beta} \vert
\ee
and therefore 
\be\label{assimpto}
\CN_n   \sim e^{n h(T)}.
\ee
{\it This result states that the number of points
on the periodic trajectories of the period n exponentially
grows  with entropy}.

Let us now define the number of irreducible  periodic trajectories (IPT)  of the period  $n$  by $\pi(n)$. The reason to 
introduce the concept of "irreducible periodic trajectories" is connected with the fact that $\CN_n$
counts not only trajectories of period $n$, but also periodic trajectories of 
a smaller period which is a devisor of $n$ as in the following example: $T^n w=T^{l_2}(T^{l_1}w)$, where $n=l_1 l_2$ and $T^{l_i}w =w$.
Therefore  the number of all points $\CN_n$ on the
periodic trajectories of period $n$  can be written in the following form:
\be\label{Npoints}
\CN_n = \sum_{l~ divi~ n  } l ~\pi(l) = n \pi(n) + l_1 \pi(l_1) + ...
+l_r\pi(l_r),
\ee
where $l_i$ divides $n$.
Excluding these reducible periodic trajectories, 
which have periods that divide $n$, one can get the number of irreducible 
periodic trajectories of period  $n$ (within $n$ points of such trajectories 
there are no points with identical coordinates). 
Thus if one represents  the $\pi(n)$ in the form
\be\label{pertraj1}
\pi(n) =
 {  \CN_n - l_1 \pi(l_1) - ...
 -l_r\pi(l_r)   \over n}
\ee
it follows from  (\ref{assimpto}) that
\be\label{pertraj}
\pi(n) \sim { e^{n h(T)} \over n}  \big( 1   -
{ l_1 \pi(l_1) + ...
 +l_r\pi(l_r) \over \CN_n } \big)
 \sim { e^{n h(T)} \over n}.
\ee
This result tells that a
system with larger entropy
\be\label{growth}
\Delta h = h(T_1) - h(T_2) >0
\ee
is more densely populated by the
periodic trajectories of the same period.

A particular C-system which we shall consider in this article was
considered  in \cite{yer1986a,konstantin,Savvidy:2015ida,Savvidy:2015jva}
 and is defined as a matrix of the 
size $N \times N$ for all $N \geq 3$:
\be
\label{eqmatrix}
T =  
   \begin{pmatrix}  
      1 & 1 & 1 & 1 & ... &1& 1 \\
      1 & 2 & 1 & 1 & ... &1& 1 \\
      1 & 3+s & 2 & 1 & ... &1& 1 \\
      1 & 4 & 3 & 2 &   ... &1& 1 \\
      &&&...&&&\\
      1 & N & N-1 &  N-2 & ... & 3 & 2
   \end{pmatrix} .
\ee
The integer matrix $T$ is of the size $N \times N$ and has 
determinant equal to one. It is defined recursively, 
since the matrix of size $N+1$ contains in it the matrix of the size $N$.  
The only variable entry in the matrix is $T_{32} =3+s$, where  $s$  is a 
free parameter which should be chosen so that eigenvalues lying on a unit circle would be avoided. It is referred in the article  as  a "special parameter".  
In this article we shall consider systems with 
$s=-1$ \cite{konstantin}.  
In order to generate trajectories one should execute the iteration  
$w_n = T^n w$ 
by choosing the initial vector $w=(w_{1},...,w_{N})$, called the ``seed",  
 with at least one non-zero component to avoid the fixed point of $T$.

{\it The asymptotic  formulas (\ref{assimpto}) and (\ref{pertraj})
define the leading behaviour of the distribution functions $\CN_n$ and $\pi(n)$,  and we are interested in finding out the behaviour 
of these functions in the next to leading order}. For that one should develop a convenient computational technique to count number of 
points on periodic trajectories. That would require efficient algorithms 
counting solutions of the equation (\ref{periodictrajectories1})
and recursive solution of  the  equations (\ref{Npoints}) and (\ref{pertraj1}).

In this article we shall explore the C-operators $T(N)$  calculating the
distribution of its periodic trajectories $\CN_n$ and $\pi(n)$ 
for the increasing values of $N$. The periodic trajectories of period $n$ are 
distributed on a rational lattice $\CL_p$ and in the second section 
we are studying how the 
size $p$ of the lattice $\CL_p$ grows with period $n$. 
The distributions functions $\CN_n$ and $\pi(n)$  are well approximated by the 
entropy dependent exponential $e^{n h}$, 
however, the difference between them is less known. In the third and forth section we are investigating the difference between them deriving 
general asymptotic formulas in next to leading order.

\section{\it Counting Periodic Trajectories}

We are interested in finding out the behaviour of $\CN_n$ and $\pi(n)$ as a function of period length $n$.
Notice that both distributions,  $\CN_n= \CN_n(T)$ and $\pi(n)=\pi(n,T)$, 
depend  on the operator $T$ of the 
dynamical system. The
asymptotic behaviour of these functions is known and is given by formulas (\ref{assimpto}) and (\ref{pertraj}). But it is not known how
fast they approach the asymptotics. We shall use numerical and analytical calculations to estimate
the rate at which they are converging. Numerically we can calculate these functions exactly up to large periods $n$ and
then compare the results  with the asymptotic behaviour. That will allow 
to study the deviation of asymptotics from the actual values.

Let us first present examples of periodic trajectories and their distribution
for the lowest operator of dimension $N=2$:
\be
\label{eqmatrix2}
T =  
   \begin{pmatrix}  
      2 & 1   \\
      1 & 1
       \end{pmatrix} .
\ee
On rational  lattice  $\CL_5$  with coordinates $\{a_1/5,a_2/5 \}$ where $a_i =0,1,...,4$ there are two trajectories of period two $\pi(2) =2$  and two trajectories of period ten $\pi(10) =2$ . Together they cover all points of the lattice $\CL_5$  with $2 \pi(2)+10 \pi(10)=24$ points, except of the origin. 
The distribution of periodic trajectories on rational lattices of size 
$p=5,4,15,11$ is presented in Table  \ref{tbl:traj}.
\begin{table}[htbp]
   \centering
   \begin{tabular}{@{} cccc @{}} 
         \toprule
      p & n & $\text{number of trajectories} $& $p^2 -1$ \\
      \midrule
        5  & \bf{2}  & 2 &  \\
	5&10&  2 & 24  \\
	 \midrule
	4 & \bf{3}   & 5 &  15\\
	 \midrule
	15 & \bf{4} &  10 &\\
	15 & 2 &  2& \\
	15 & 10 &  2& \\
	15 & 20 &  8& 224\\
	 \midrule
	11 & \bf{5} &  24& 120\\
      \bottomrule
      \end{tabular}
   \caption{The number of periodic trajectories of period $n$ on a lattice of the size $p$. The trajectories 
   completely cover all points $p^2-1$ of the rational lattice $\CL_p$.  The lattices 
   $p=5,4,15,11$ appear in sequence as we solve the 
   equation (\ref{inverse}) for the
   periods $n=2,3,4,5$. {\it All irreducible trajectories of period two 
   first appear on lattice $p=5$, of period three on $p=4$, of period four on $p=15$ and of period five on $p=11$}.}
   \label{tbl:traj}
\end{table}

In our first approach we shall calculate the distributions
$\CN_n$ and $\pi(n)$ based on the  equation  (\ref{periodictrajectories1}) which defines the 
coordinates of the periodic solutions
and use the formulas (\ref{Npoints}) and (\ref{pertraj1}) to extract $\CN_n$ and $\pi(n)$. The solutions of the equation (\ref{periodictrajectories1})
can be written in the form  :
\be\label{inverse}
w = (T^n-I)^{-1}b,~~~~(mod ~1),
\ee
which expresses the coordinates of the periodic trajectories in terms of an integer vector $b$. Choosing a period $n$ and varying the components of 
the integer vector $b$ in sufficiently large intervals, one can compute all 
coordinates of points that are
belonging to different trajectories of the same period $n$.
The inverse matrix $(T^n-I)^{-1}$ in (\ref{inverse})  is not integer because its elements are equal to the minors divided by the determinant of the matrix $T^n-I$, which is not equal to one. We shall denote the corresponding maximum denominator by $p$, so that all elements of the
matrix $(T^n-I)^{-1}_{i,j}$ have the simplest form $M_{i,j}/p$
(meaning that 
it cannot be reduced further). It follows then that all coordinates $w$
of the periodic trajectories will be of the form $\{a_1/p,...., a_N/p\}, i=1,...N$ and therefore will belong to a rational lattice $\CL_p$.   
{\it That means that the value $p$ defines a rational lattice $\CL_p$ on which all trajectories of period $n$ can be settled.}
\begin{figure}[h]
    \centering
    \begin{subfigure}[h]{0.45\textwidth}
        \includegraphics[width=1\textwidth]{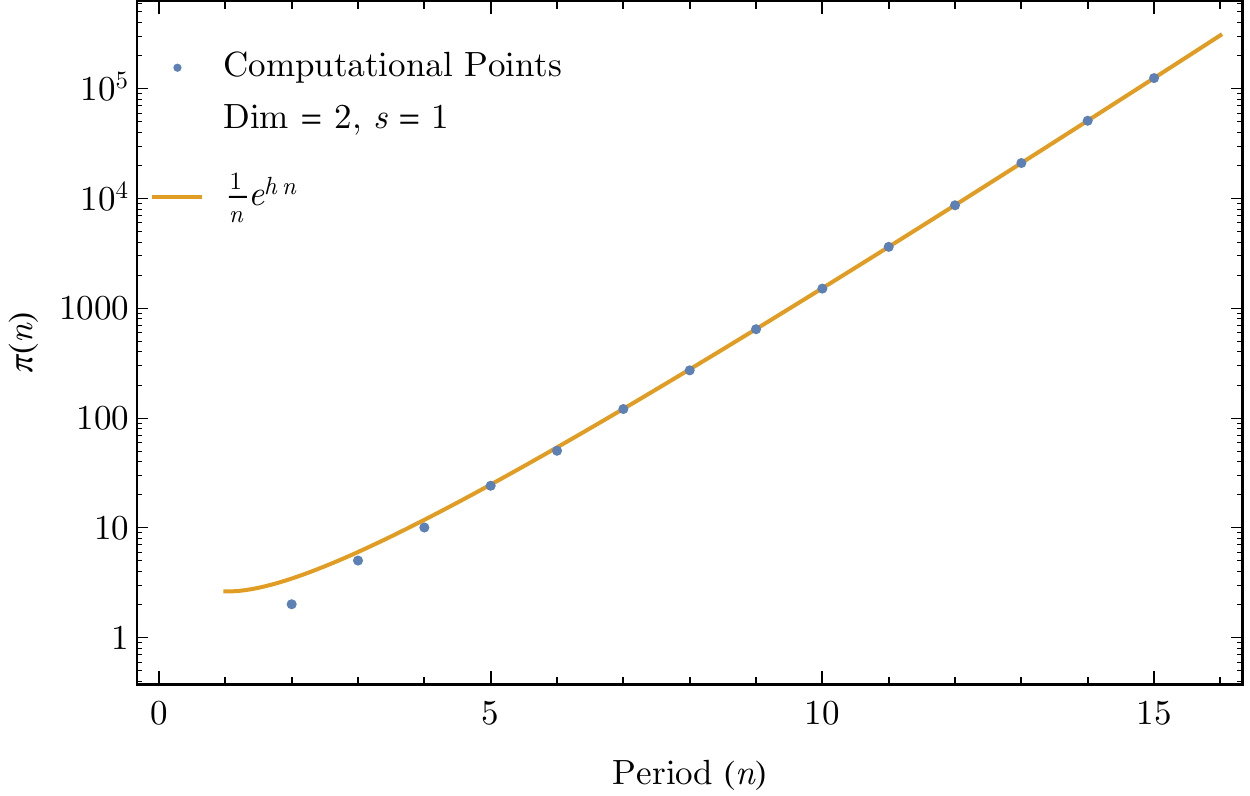}
    \end{subfigure}
   \centering
    \begin{subfigure}[h]{0.45\textwidth}
        \includegraphics[width=1\textwidth]{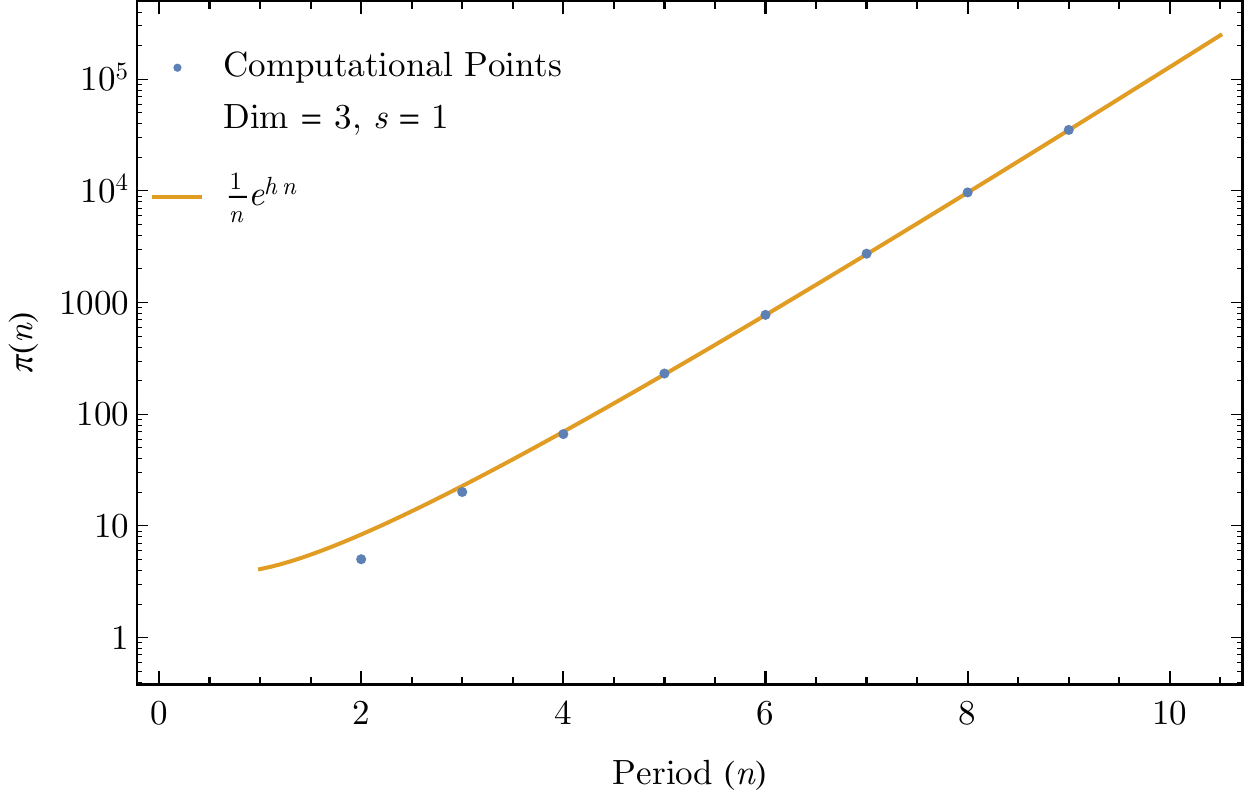}
    \end{subfigure}
    \caption{The number of periodic trajectories $\pi(n)$ for period
    $n$ for the matrices of the size $N=2$ and $N=3$. The presented data is for $N=2$, $n \leq15$ and $N=3$, $n \leq 9$ respectively. The graphs are given in a logarithmic scale.
    As one can see $\pi(n)$ very closely
    follows its asymptotic behaviour $\pi(n)
 \sim  e^{n h(T)} / n $~  even for small periods $n$. }
    \label{fig1}
\end{figure}
The question  is: in which interval the integer values of $b$ should 
variate in order to generate all points of the periodic trajectories of a given period $n$, but without 
duplication of the solutions? One can easily restrict a required  interval by taking into consideration the properties of modulo operation. 
As we mentioned before, the elements of the inverse matrix that act on $b$ are not integer, but rational numbers with maximum denominator $p$. For components of $b$ bigger or equal $p$ the modulo operation will generate trajectories with coordinates in the interval  $[0,1)$, thus it is sufficient for the coordinates of vector $b$ 
to variate within the interval $[0, p-1]$.  
The problem is that the identical solutions can still
appear many times  and 
therefore one should  further restrict the interval in which 
components of $b$ 
are varying in order to exclude identical 
solutions. We have developed an algorithm which 
excludes identical solutions and the results are presented on  Fig.\ref{fig1} and Fig.\ref{fig2}.

\begin{figure}[h]
    \centering
    \begin{subfigure}[h]{0.48\textwidth}
        \includegraphics[width=1\textwidth]{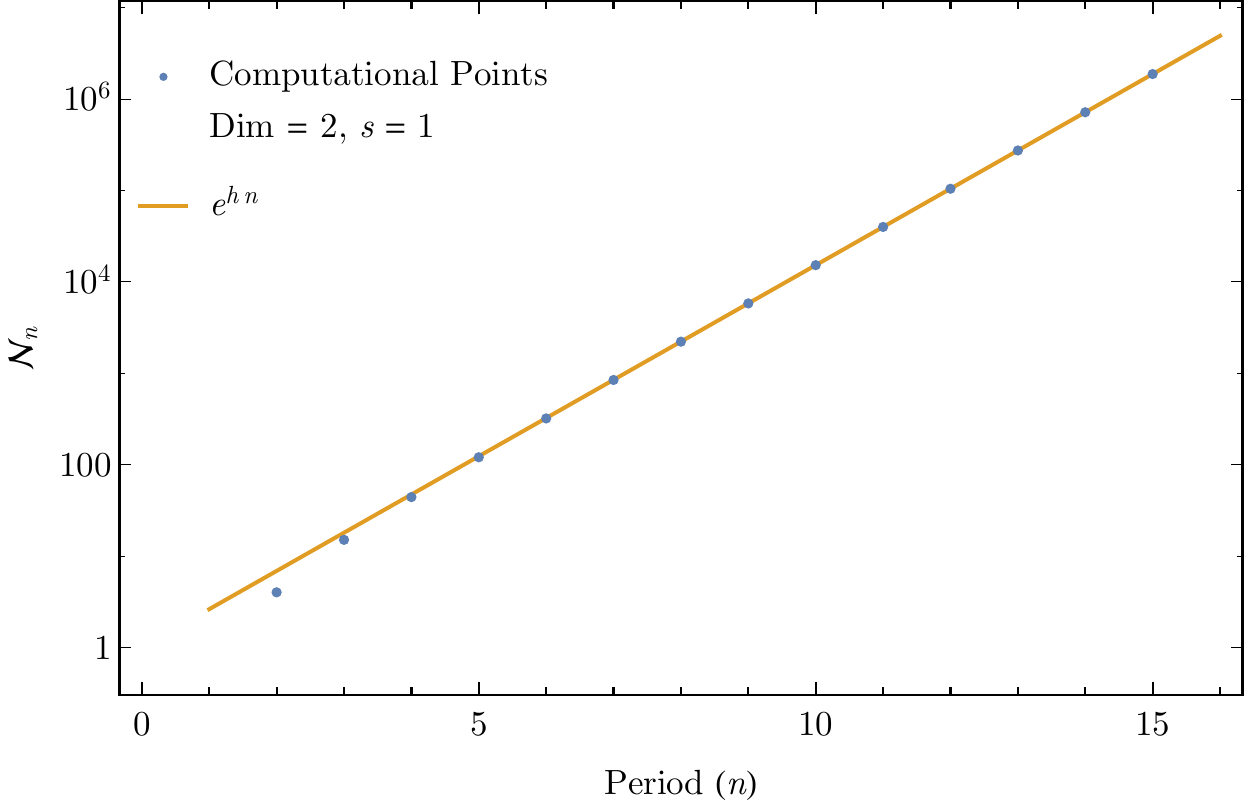}
    \end{subfigure}
    \centering
    \begin{subfigure}[h]{0.48\textwidth}
        \includegraphics[width=1\textwidth]{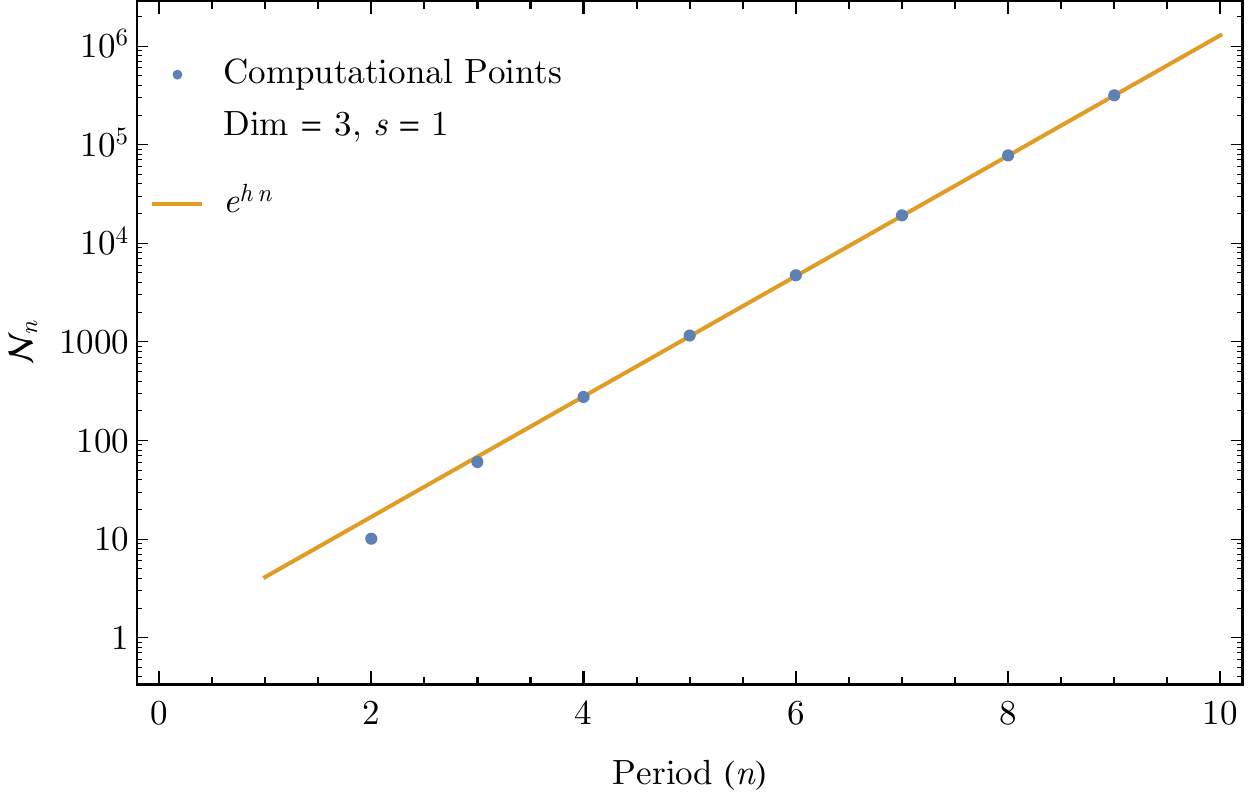}
    \end{subfigure}
    \caption{The number of points $\CN_n$ on trajectories of period $n$
    given in a logarithmic scale for a matrix of dimension $Dim=N$. The presented data is for $N=2$, $n \leq 15$ and $N=3$, $n \leq 9$ respectively.
    As one can see, $\CN_n$ very closely
    follows its asymptotic behaviour $\CN_n
 \sim  e^{n h(T)} $~  even for small periods $n$.}
    \label{fig2}
\end{figure}

\begin{figure}[h]
    \centering
    \begin{subfigure}[h]{0.48\textwidth}
        \includegraphics[width=1\textwidth]{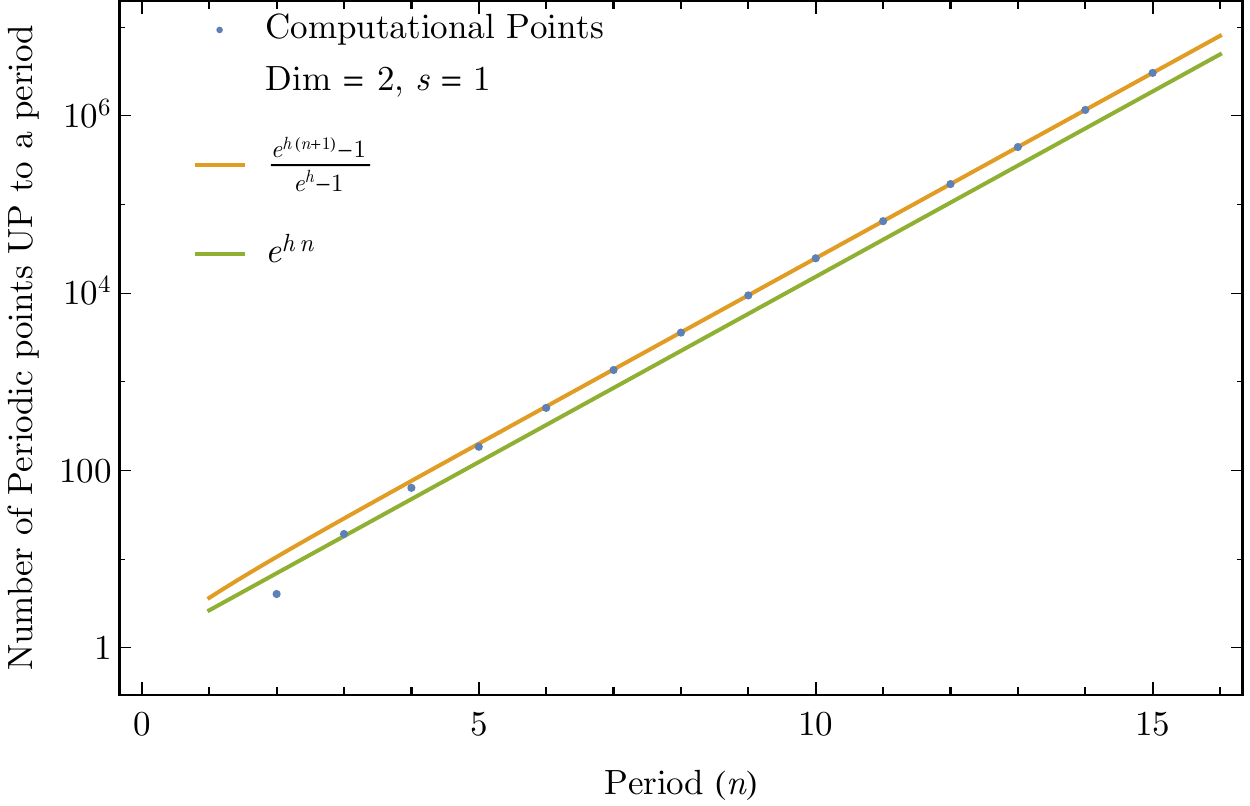}
    \end{subfigure}
    \centering
    \begin{subfigure}[h]{0.48\textwidth}
        \includegraphics[width=1\textwidth]{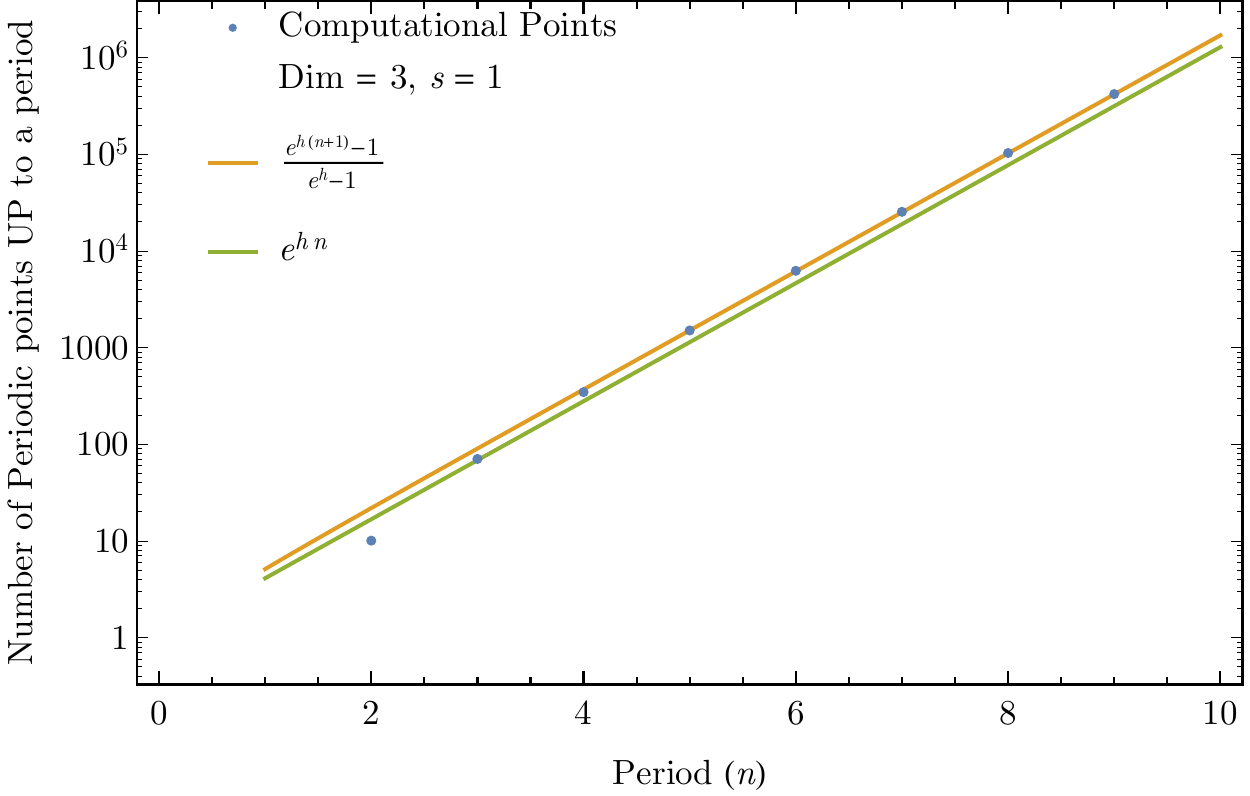}
    \end{subfigure}
    \caption{The number of \emph{total}  points $\sum_i^n \CN_i$ on the trajectories of periods less or equal to $n$ for the matrix of size $N=2$ and $N=3$. The presented data is for $N=2$, $n=15$ and $N=3$, $n=9$ respectively. It also shows the $\pi(n)$ (green line) and $\CN_n$ (red line) as well.}
    \label{fig3}
\end{figure}

The figures Fig.\ref{fig1} and Fig.\ref{fig2} represent the distribution functions $\CN_n$ and $\pi(n)$ for matrices of the size $N=2,3$ and special parameter $s=-1$. As one can see from the graphs, the distributions $\CN_n$ and $\pi(n)$ are well approximated by the asymptotic formulas even for small values of $n$.
As expected from the formula (\ref{growth}), the number of periodic trajectories of a given period rises faster for a system with a larger  entropy. For example, the MIXMAX matrix of the size $N=2$ has $640$ trajectories with period $n=9$, while the matrix of dimension $N=3$ has $34892$ trajectories of the same period! In Fig.\ref{fig3} we have plotted the sum of all points of periodic trajectories up to a given period, it is a sum of a geometric series $(1-r^{n+1})/(1-r)$ with $r = e^h$.

For a more efficient calculation of distributions  $\CN_n$ and $\pi(n)$ we 
shall use the representation of  $\CN_n$ in terms of determinant, which is given by equation (\ref{numbers}). This way we are getting less information about actual structure of the trajectories, about their 
coordinates, but it allows to proceed counting trajectories 
for much larger periods.  Indeed, it is relatively easier 
 to calculate the determinants of large powers of a given matrix and therefore 
 one can study $\CN_n$ for fairly large periods.
It is important to notice again that $\CN_n$ counts the number of \emph{all} points on reducible and irreducible periodic trajectories that have a period $n$. For example, $\CN_{10}$ includes the points that have period exactly equal to $10$ as well as the points on the trajectories which have the periods  which are dividers of $10$, like
$\CN_2=2~ \pi(2)$, $\CN_5=5~ \pi(5)$, therefore in this case we have
$$
\CN_{10} = 10~ \pi(10) + 5 ~\pi(5) + 2~ \pi(2) = 15124 = 10 \times 1500 + 5 \times 24 + 2\times 2.
$$
This formula also demonstrates that to get the number of irreducible trajectories 
of period ten $\pi(10)$, one should subtract from $\CN_{10}$ the contribution of reducible trajectories with periods which are dividers of $10$. 
So the formula would be:
$$
\pi(10)=(\CN_{10}-5 \pi(5)-2 \pi(2))/10.
$$
{\it For prime periods $n$ the relation has a fundamental form }
\be
\CN_n = n~\pi(n)~.
\ee
The knowledge of $\pi(n)$ for prime $n$ allows to find out distribution function $\pi(n)$ for non prime values of $n$, as in the above case, one should know the values of $\pi(2)$ and $\pi(5)$ to get $\pi(10)$. This can be done algorithmically, in sequence, for increasing prime values of $n$. Thus, one should calculate the determinant of the matrix $T^n-I$,  then subtract the number of periodic trajectories that have periods that divide $n$, as in the example considered above. This method is much faster and allows to study the distribution
functions for matrices of larger sizes $N$.

\begin{figure}[h]
    \centering
    \begin{subfigure}[h]{0.48\textwidth}
        \includegraphics[width=1\textwidth]{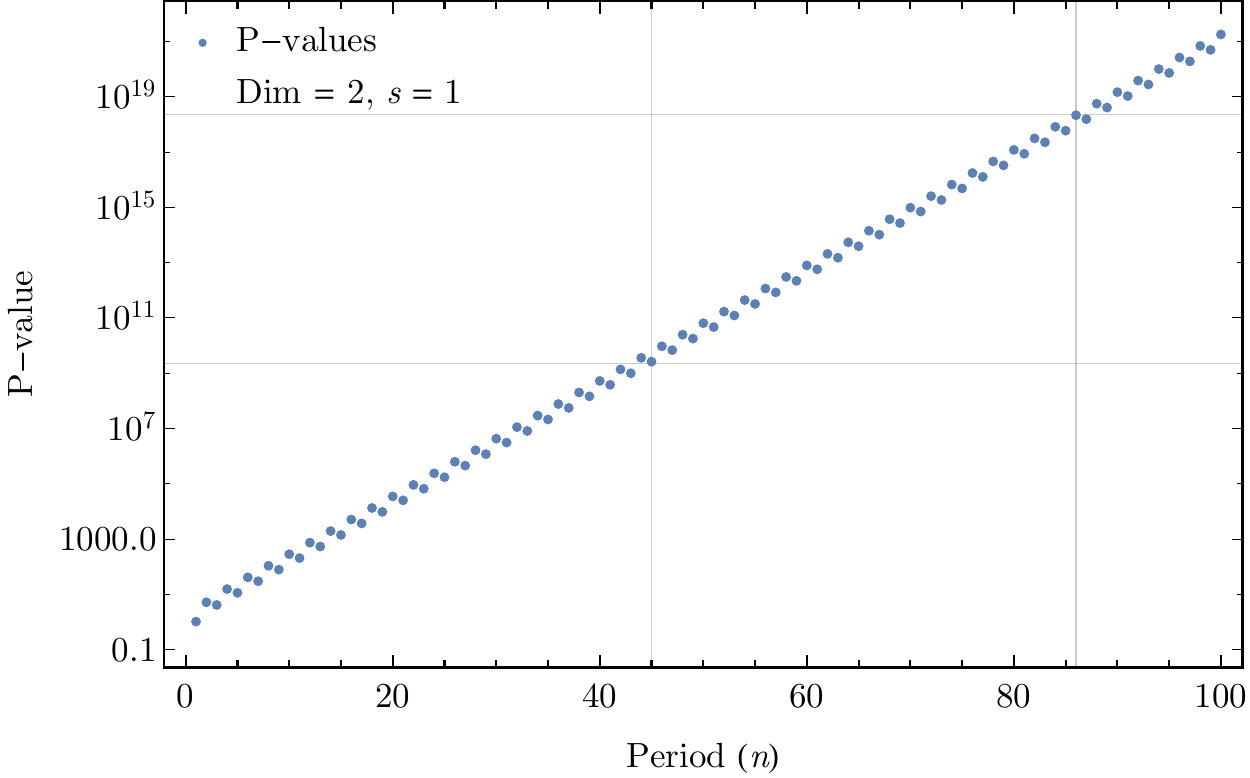}
    \end{subfigure}
    \centering
    \begin{subfigure}[h]{0.48\textwidth}
        \includegraphics[width=1\textwidth]{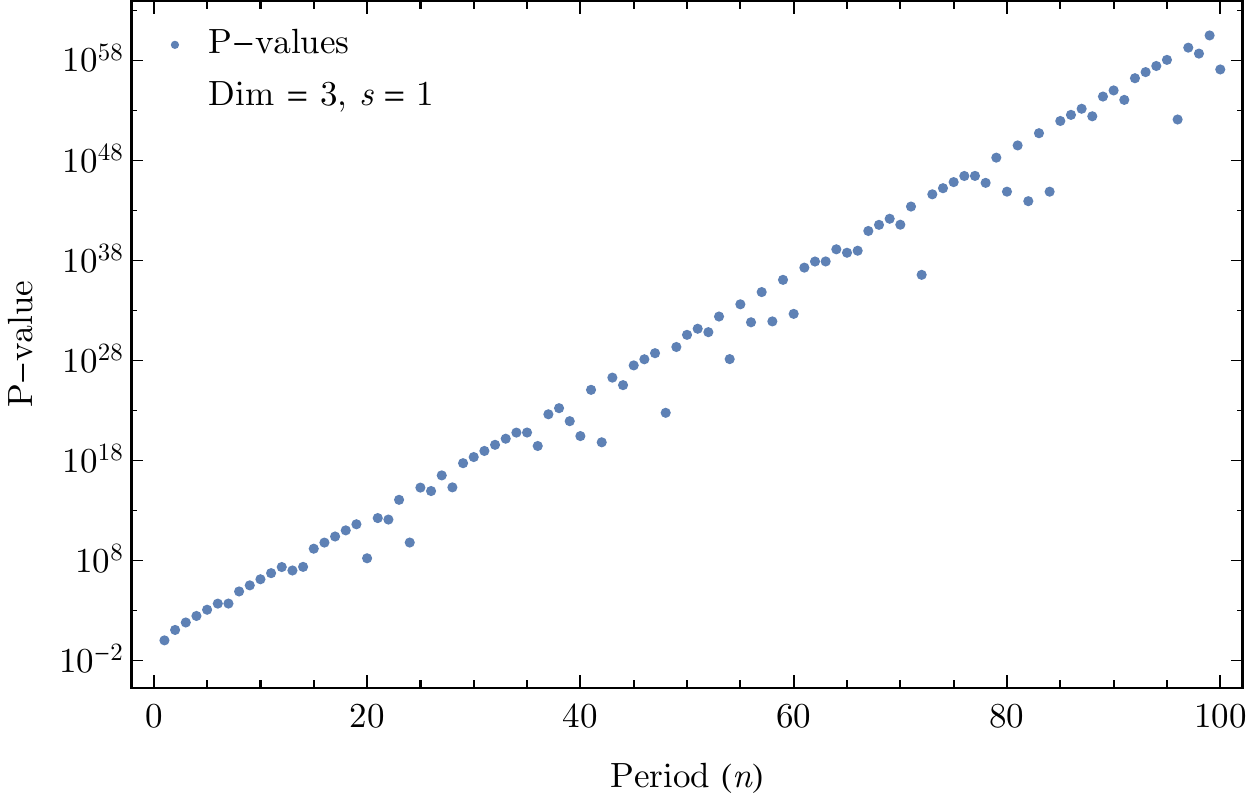}
    \end{subfigure}
    \caption{The graphs show how the lattice size $p$  grows as a function of period $n$ for the matrices
    of size $N=2$ and $N=3$. The lattice size $p$  grows exponentially with period $n$ and represent the size of the lattice that includes \emph{all} periodic trajectories of a given period $n$. The lower points on the graph describe periods for which the determinant of the system $Det(T^n-I)$ and the minors of the matrix $(T^n-I)$ have a common divider.}
    \label{fig4}
\end{figure}

The Fig.\ref{fig6} presents the results for the matrices of the size $N=2,3,17,256$ and of special parameter $s=-1$.  In agreement with our previous results the distributions $\CN_n$ and $\pi(n)$ are well approximated by the asymptotic formulas even for small values of $n$. For large matrices 
$N=17,256$, the corresponding entropies are much larger and the exponential term prevails. This is the reason why the functions $\CN_n$ and $\pi(n)$ on the graph seem to coincide.

What is surprising here is that the size $p$  of the rational lattice $\CL_p$ grows exponentially with period $n$. Indeed, inspecting the  Fig.\ref{fig4} 
one can see that {\it all } trajectories of a relatively small period $n = 86$ are distributed on a rational lattice $\CL_p$ of a huge 
size $p=2^{61}-1$.
This effect seems to contradict to the statement that 
even very large periods can be realised on much smaller rational lattices. The MIXMAX generator is 
using a lattice of the size $p=2^{61}-1$ and has periods of order $n = (p^N -1)/(p-1)$ \cite{konstantin}.
The resolution of this paradox lies in the fact that  in reality {\it some of the 
periodic trajectories of period $n = 86$  are settled on a coarser 
sublattices $\CL_p^{'} \in \CL_p$}.
In particular, for the matrix of the size $N=2$, all $\pi(10)=1500$ trajectories of period $n=10$ are on the rational lattice of the size $p=275$, and the question is: Do all these trajectories lie on a such fine lattice or some of them are on much coarser sublattices? Indeed, as one can 
see in Table \ref{tbl:traj},  two trajectories of period $n=10$ appear on a  sublattice of the size  $p=5 $. The reason why this happens is that the denominator $p$ is decomposable into small primes: $p=275=5\cdot5\cdot11$ and for the vectors $b$ of the form $b = a\cdot5 \cdot11  $ the coordinates of the trajectory $w=(T^{10}-I)^{-1} b =  { 1 \over  5}a$, where $a$ is an integer vector,  will be found on the lattice of the size $p=5$. {\it Thus the rational lattice $\CL_p$ is the 
lattice which contains $\bf{all}$  irreducible trajectories of period $n$, but 
some of them can be settled on coarser sublattices. }

\begin{figure}[h]
    \centering
    \begin{subfigure}[h]{0.48\textwidth}
        \includegraphics[width=1\textwidth]{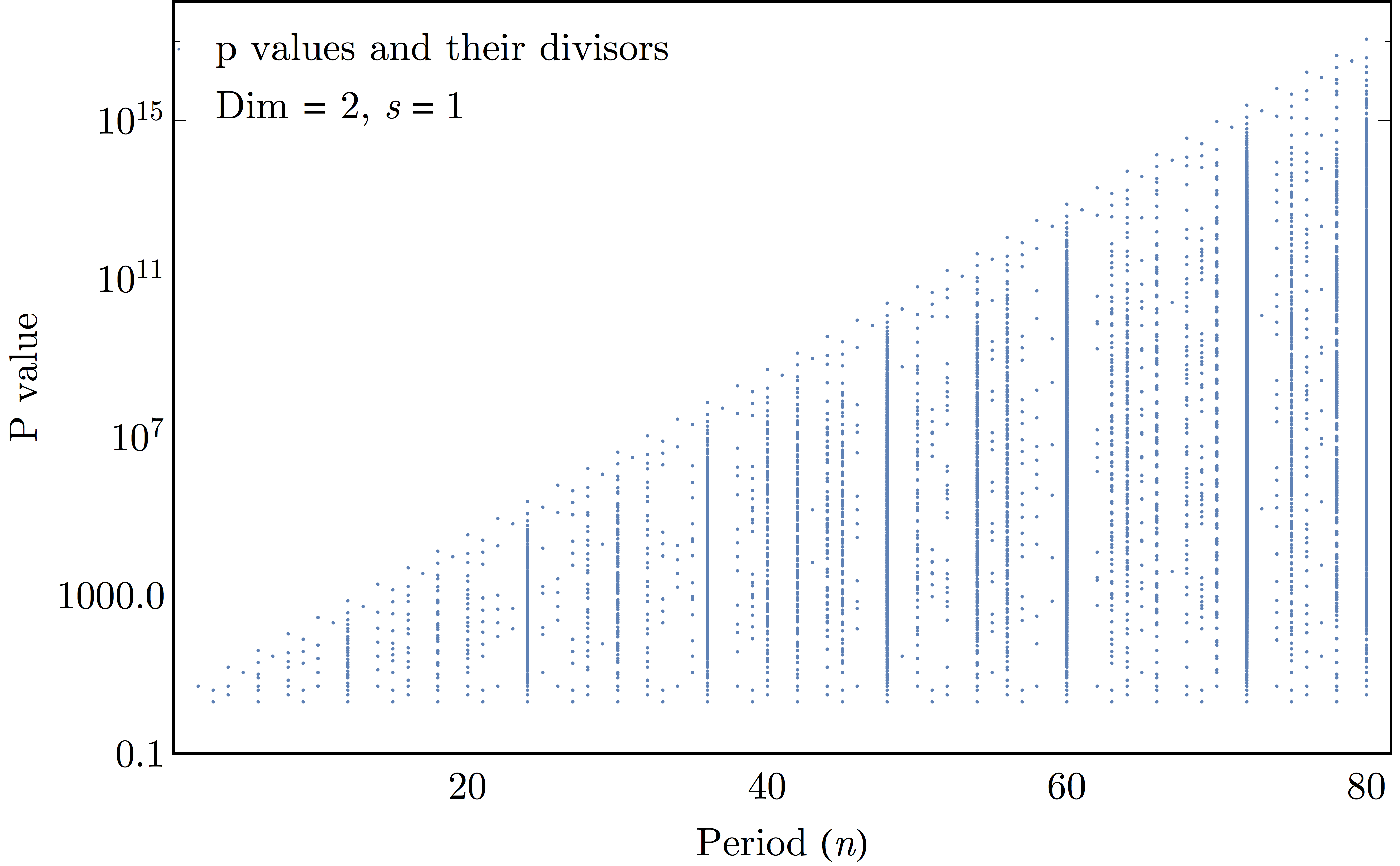}
    \end{subfigure}
    \centering
    \begin{subfigure}[h]{0.48\textwidth}
        \includegraphics[width=1\textwidth]{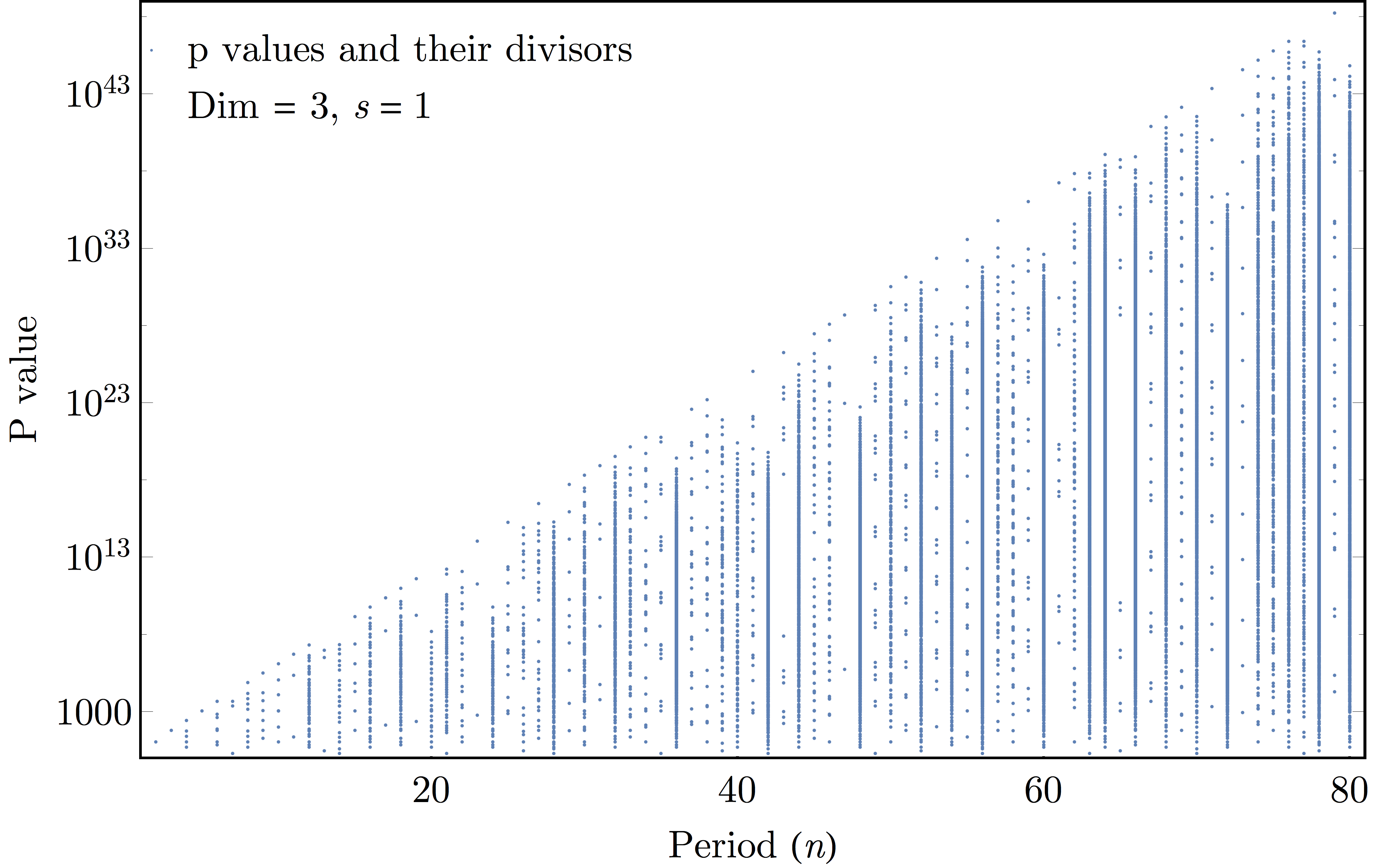}
    \end{subfigure}
    \caption{The distribution of dividers $p=p_1...p_k$ of the rational lattice $\CL_p$ for periods in the interval $n=2,..80$ and 
    matrices of the size $N=2,3$. The dots in the vertical direction show the values of $p_1, ...,p_k$.
    }
    \label{fig5}
\end{figure}

This general  property is demonstrated on Fig. \ref{fig5}, where we have plotted the devisors of the rational lattices $\CL_p$.
One can see that there are very coarse  sublattices 
of dimension equal to the devisors of $p$.  Thus 
periodic trajectories of a given period $n$ can be found on coarser  sublattices that are devisors of $p$. That is why 
one can generate pseudo-random numbers with a huge period without having to resort to extremely fine rational lattices, gaining in speed and efficiency
\cite{konstantin,Savvidy:2015ida,Savvidy:2015jva,hepforge,cern,root,geant}.
The generators with $N=240$ and $N=256$ has the best combination of speed, reasonable size of the state, and availability for implementing the parallelization and is currently available  generator in the ROOT and CLHEP software packages at CERN for scientific calculation \cite{cern,root,geant}.

\begin{figure}[h]
    \centering
    \begin{subfigure}[h]{0.49\textwidth}
        \includegraphics[width=1\textwidth]{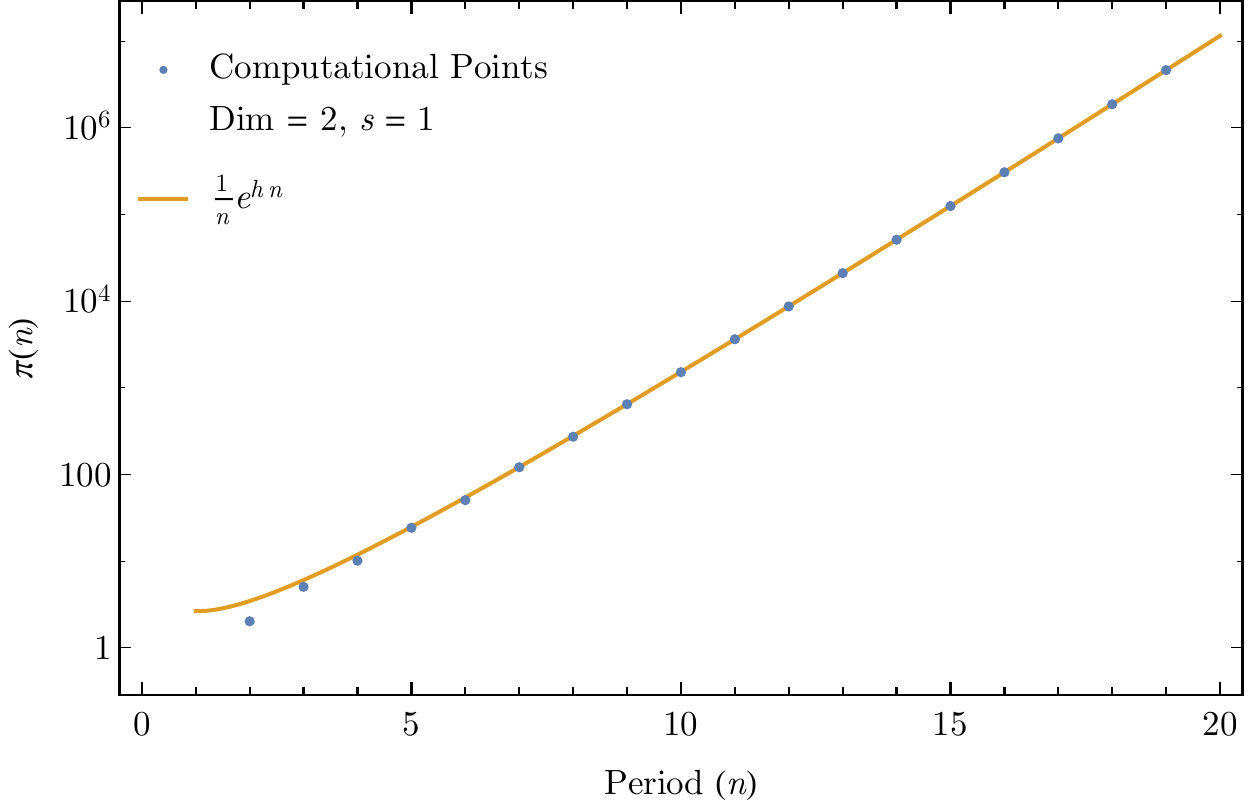}
    \end{subfigure}
    \centering
    \begin{subfigure}[h]{0.49\textwidth}
        \includegraphics[width=1\textwidth]{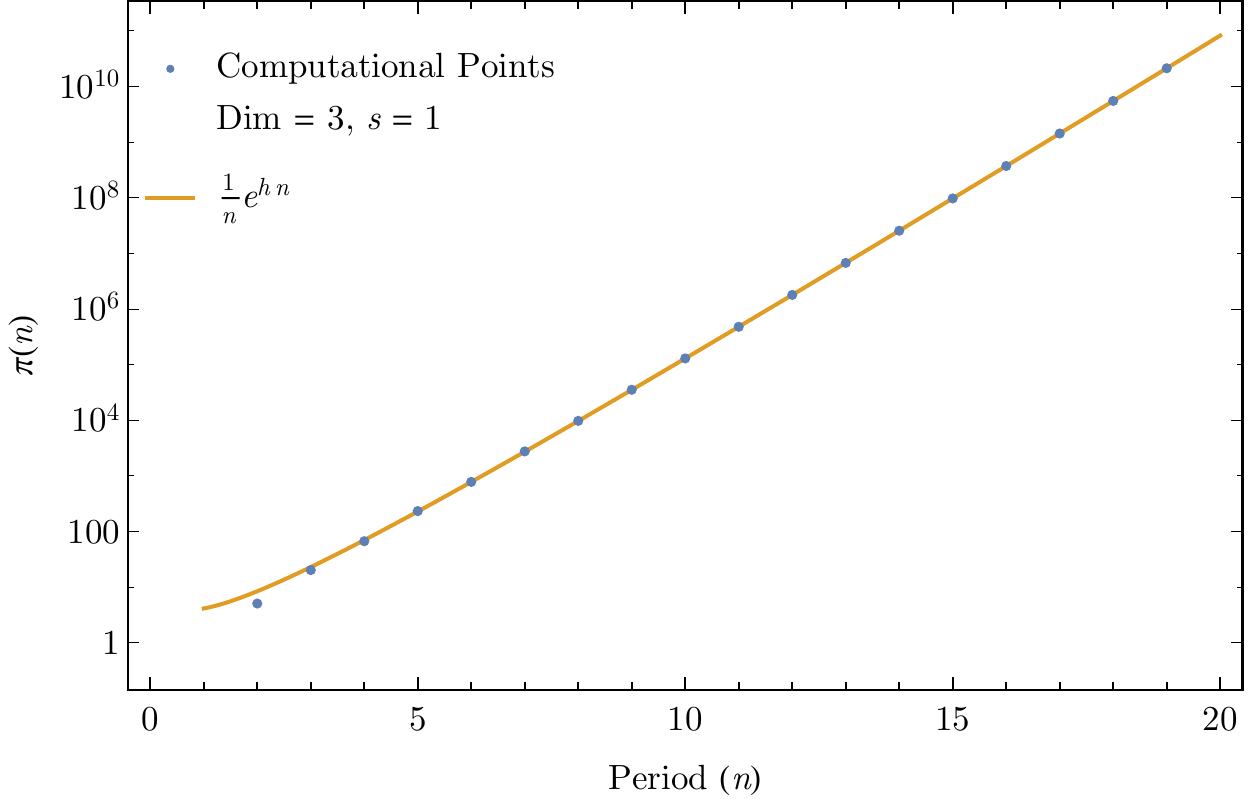}
    \end{subfigure}
    \centering
    \begin{subfigure}[h]{0.49\textwidth}
        \includegraphics[width=1\textwidth]{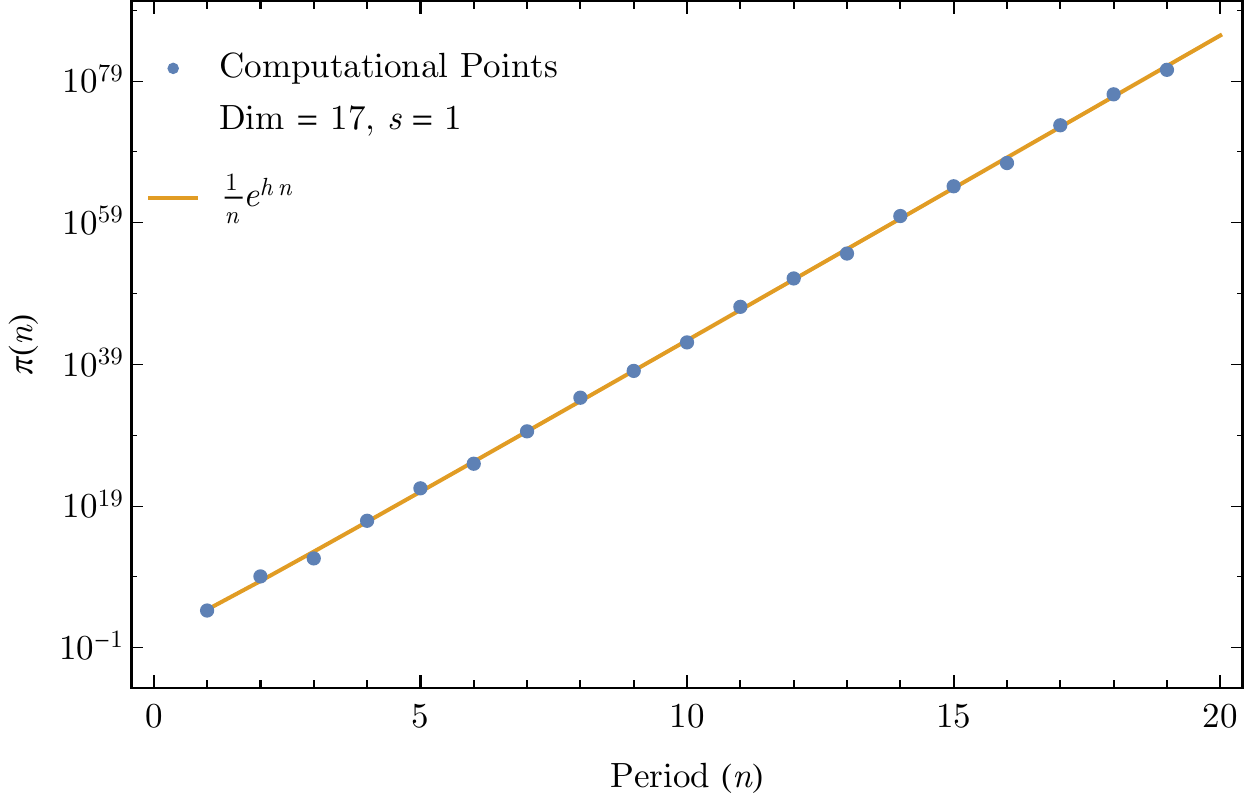}
    \end{subfigure}  
        \centering
    \begin{subfigure}[h]{0.49\textwidth}
        \includegraphics[width=1\textwidth]{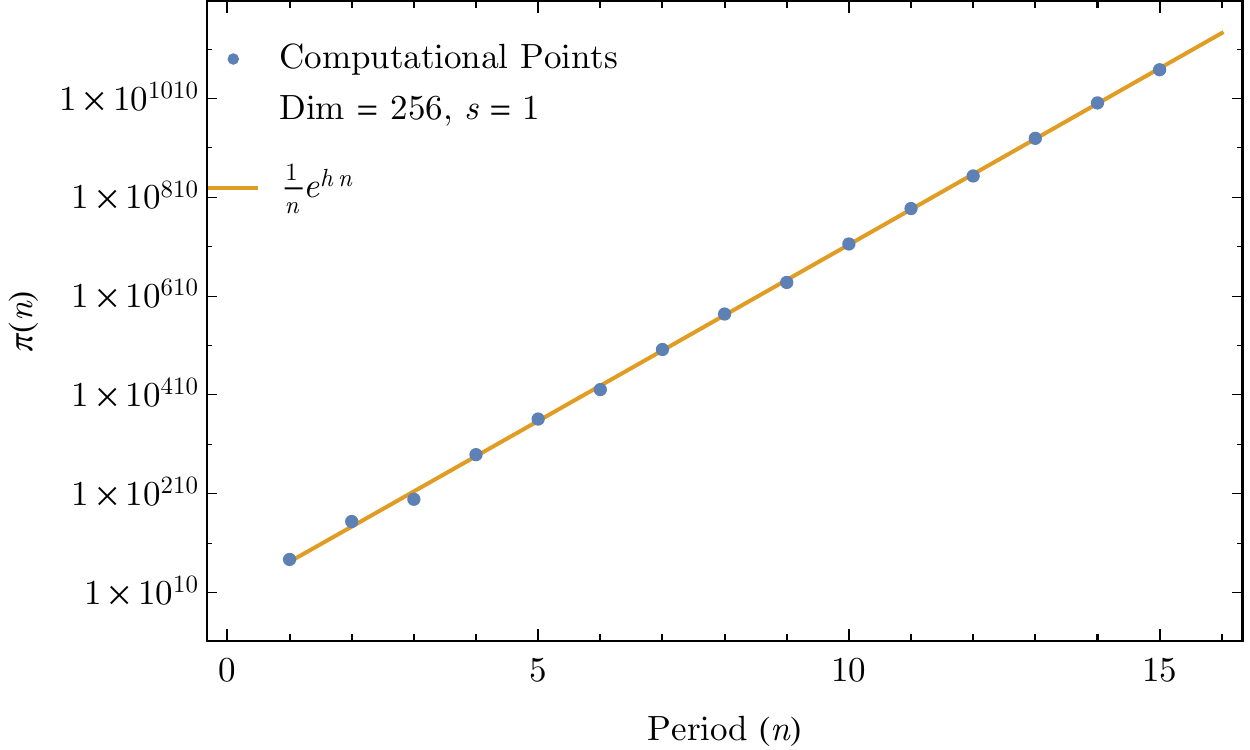}
    \end{subfigure}   
    \caption{Number of periodic trajectories $\pi(n)$ for period
    $n=2,...,20$ for matrix dimensions $N=2$, $N=3$, $N=17$ and $N=256$ in a logarithmic scale. For these graphs a more efficient calculation of distributions  $\CN_n$ and $\pi(n)$ was used 
representing  $\CN_n$ in terms of determinant, which is given by equation (\ref{numbers}). }
    \label{fig6}
\end{figure}

\section{\it Distribution of Periodic Trajectories  for $N = 2$}  

The distribution of periodic trajectories for the matrix  of the smallest size $N = 2$
can be calculated analytically because of its ultimate relation with the Fibonacci series.
For $N = 2$ the $T$ matrix is given by
\[
T = \begin{pmatrix}
2 & 1\\
1 & 1\\
\end{pmatrix} .
\]
Using the definition and properties of Fibonacci series $F_n$ we can calculate 
the powers of matrix $T$ and $\CN_n$ for odd $n$,
\[
T^n = \begin{pmatrix}
F_{2 n + 1} & F_{2 n}\\
F_{2 n} & F_{2 n - 1}\\
\end{pmatrix}, \quad
\CN_n = |\Det [T^n - 1]| = (F_{n - 1} +  F_{n + 1})^2.  
\]
Above equation shows that for $n$ odd, $\CN_n$ is a square of an integer number
which has meaningful consequences. Similarly, we can calculate the inverse matrix
\[ (T^n - 1)^{-1} =
\frac{1}{F_{n - 1} +  F_{n + 1}} \begin{pmatrix}
- F_{2 n - 1} & F_{2 n}\\
F_{2 n} & - F_{2 n + 1}\\
\end{pmatrix}.
\]
It is easy to show by induction that matrix elements cannot be further reduced and the 
lattice  size is $p = F_{n - 1} +  F_{n + 1}$ and therefore $\CN_n = p^2$.
The property that $\CN_n = p^2$ is specific to $N = 2$.
As a result all points on a lattice $\CL_p $ belong to trajectories
with period $n$ or dividers of $n$. 
If $n$ is a prime number then there are
\[ \pi(n) = \frac{p^2 - 1}{n}, \quad p = F_{n - 1} +  F_{n + 1} \]
trajectories on this lattice all of which have period $n$ and which are shared by all points except 
the origin.
Knowing $\CN_n$ the number of trajectories $\pi(n)$ with odd period $n$ can be calculated by taking out points with
different periods (which have to be dividers of $n$),
\[ \pi(n) = \frac{1}{n} \left(\CN_n - \sum_{l~ divi~ n  } l \, \pi(l) \right) = \frac{1}{n} \sum_{l~ divi~ n  } \mu(l) \CN_l, \]
where $\mu(l)$ is the M\"obius function which ensures proper cancellation of points.
This result shows that $\pi(n)$ for prime $n$ are fundamental objects 
which can be used to express arbitrary $\pi(n)$.

\section{\it Convergence of $\CN_n$ and $e^{n h}$}
\label{sec:convergence_of_cn_and_e_n_h_}

The number of all points $\CN_n$ on the periodic trajectories of the period $n$ 
converges very fast to $e^{n h}$  as can be seen on Fig.\ref{fig1}-\ref{fig3} and Fig.\ref{fig6}.
However, due to the exponential behaviour of those quantities the plots do not capture 
the real extent of the difference between them 
and the question how well $\cN_n$ is described by an exponential 
growth with the entropy remains open.
The number of points $\CN_n$ in trajectories of period $n$ is given by following formula 
(\ref{numbers} ) which can be rewritten in the form:
\begin{equation}\label{eq:Nnchi}
\cN_n = \left| \Det \ T^n - 1 \right| = \prod_i \left| \lambda_i^n - 1 \right| = e^{n h} \prod_i (1 - \chi_i^n), 
\end{equation}
where
\[ 
\chi_i = \left\{
	\begin{array}{rl}
		\lambda_i	& \text{if } |\lambda_i| < 1,\\
		\frac{1}{\lambda_i}	& \text{if } |\lambda_i| > 1.
	\end{array}		\right.
\]
By definition, $|\chi_i| < 1$ for all $i = 1, \dots N$.
Because in our case $T$ is a real matrix, and thus $\tr \ T^n$ are also real for all $n$,
its complex eigenvalues will always occur as conjugate pairs. 
Expanding the product in (\ref{eq:Nnchi}) we can calculate the deviation $\CN_n - e^{n h}$
and its approximations as a simple function of quantities $\chi_i$,   
\begin{align}\label{eq:diffchi}
\Delta = \CN_n - e^{n h} &= e^{n h} \left(- \sum_i \chi_i^n + \sum_{i < j} \chi_i^n \chi_j^n - \cdots\right) = \Delta_1 + \Delta_2 + \cdots,\nonumber \\
\Delta_1 &= - e^{n h} \sum_i \chi_i^n, \quad \Delta_2 = \frac{1}{2}e^{n h} \sum_{i \neq j} \chi_i^n \chi_j^n, \quad \dots .
\end{align}
The above expansion terms  $\Delta_k$ are symmetric polynomials of $\chi_i$'s and,
while complex eigenvalues appear in complex conjugate pairs, they are also real.
The first order approximation $\Delta_1$ of the difference in (\ref{eq:diffchi}) is accurate 
for sufficiently large $n$ and is itself dominated by the eigenvalue or eigenvalues with largest $|\chi_i|$.
Therefore, the sum in expression defining $\Delta_1$ decays exponentially with $n$ as $\max_i |\chi_i|^n$ 
and the deviation $\Delta$ behaves as 
\begin{equation}\label{eq:hprim}
\cN_n - e^{n h} = c(n) e^{n h'}, \quad  h' = h + \ln \max_i |\chi_i| < h.
\end{equation}
The coefficient $h'$ is determined by an eigenvalue of $T$ closest to the unit circle in complex plane
and $c(n)$ is the pre-exponential factor.
Our aim is to calculate the exponential coefficient $h^{'}$ in (\ref{eq:hprim}) and the pre-factor $c(n)$.

Let us consider a function $\tr \ \frac{1}{z - B}$ which is a meromorphic function with poles 
being the eigenvalues of matrix $B$.
From Cauchy's theorem it follows that a contour integral can be used to calculate
a sum over eigenvalues in some region $D$ of function $f(z)$,
\[ \sum_{\lambda \in D} f(\lambda) = \frac{1}{2\pi i} \tr \oint_{\partial D} \frac{f(z)}{z - B} \dd z \]
as long as $f(x)$ is an analytic function in a region enclosed by loop $\partial D$ with no zero'es coinciding with eigenvalues of $B$.
A complex integral over a unit circle as a contour can be used to
sum powers of $\lambda_i$ and $\lambda_i^{-1}$ whose modulus is less than one and calculate the first order approximation $\Delta_1$,
\begin{align}\label{eq:trunit}
\Delta_1& = - e^{n h}  \sum_i \chi_i^n  = -\frac{e^{n h}}{2\pi i} \tr \oint_{C(0, 1)} \left[\frac{z^n}{z - T} + \frac{z^n}{z - T^{-1}}\right] \dd z=  \nonumber\\
&=  - \frac{e^{n h}}{2\pi i} \ \tr \oint_{C(0, 1)} \frac{(2 z - B) z^n}{z^2 - z B + 1} \dd z, \quad B = T + T^{-1} .
\end{align}
Equation (\ref{eq:trunit}) provides a compact formula for the first order approximation of $ \cN_n - e^{n h}$ defined in (\ref{eq:diffchi}).

\vspace{1em}
\noindent {\it Dimension $N = 2$}
\vspace{1em}

\begin{figure}
\begin{center}
\includegraphics[width=0.7\linewidth]{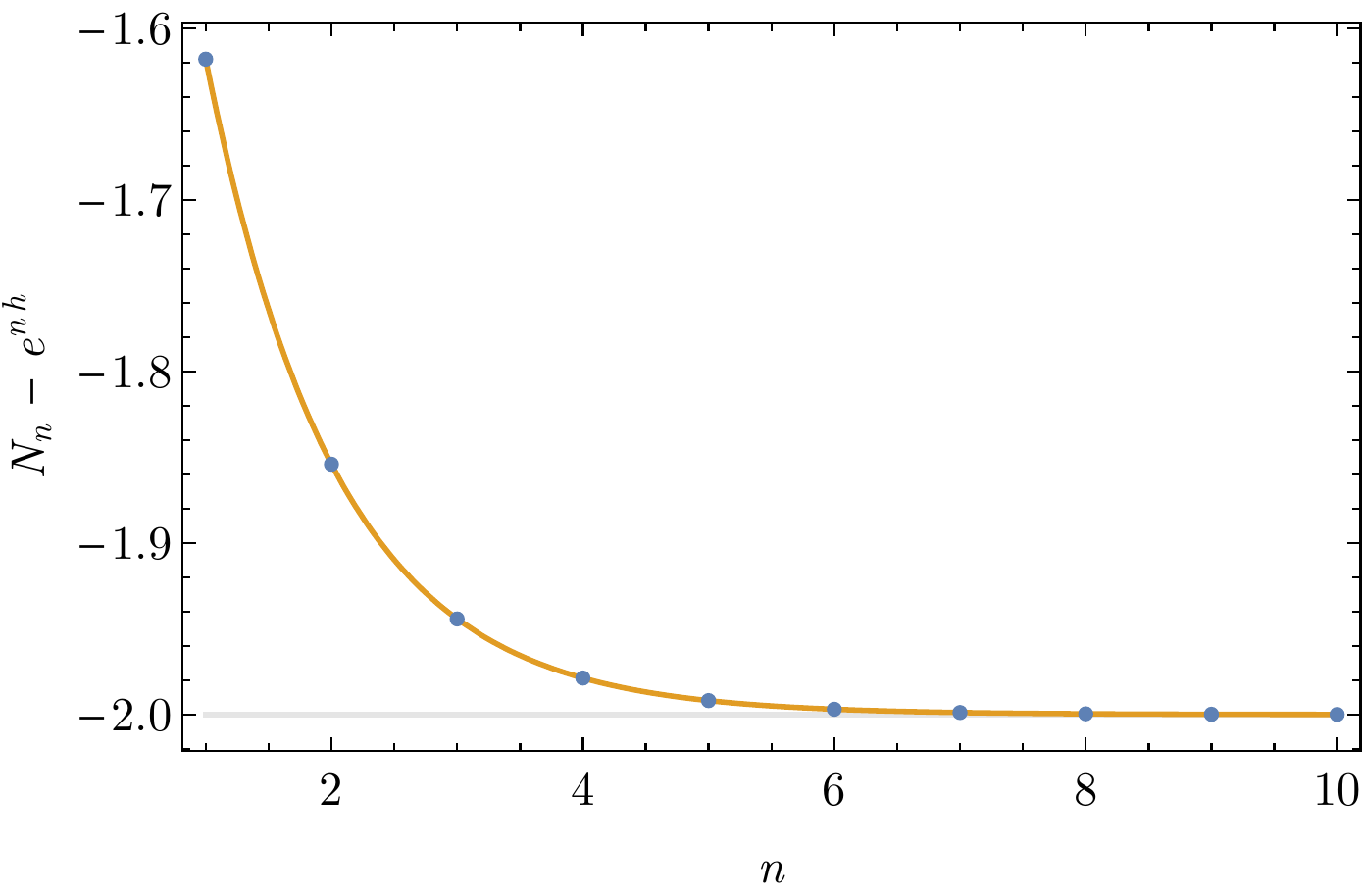}	
\caption{Difference between the number of all points $\cN_n$ on the periodic trajectories 
of the period $n$ and its estimate exponential growth with entropy $e^{n h}$ for $N = 2$.}
\label{fig:diff2}	
\end{center}
\end{figure}

To verify expressions (\ref{eq:diffchi}) and (\ref{eq:hprim}) we start with the simplest case.
For $N = 2$ the matrix $T$ has two positive eigenvalues
$ 
\lambda_1 = \left( \frac{3+\sqrt{5}}{2} \right), \quad  \lambda_2 = \lambda_1^{-1} = 
\left( \frac{3-\sqrt{5}}{2} \right),
$
and the entropy $h = \ln \lambda_1$.
The difference $\Delta$ can be calculated directly,
\[ \cN_n - e^{n h} = (\lambda_1^n - 1)(1 - \lambda_2^n) -  \lambda_1^n = - 2 \left(1 - \frac{1}{2} e^{-n h}\right) \underset{n \to \infty}{\longrightarrow} -2 .\]
The difference converges exponentially to a constant as shown in Fig. \ref{fig:diff2}, thus the relative difference 
$\left|\frac{\cN_n - e^{n h}}{e^{n h}}\right|$ decays exponentially.
Blue dots were obtained by counting the number of points on periodic  trajectories and calculating the difference.
The function $- 2 \left(1 - \frac{1}{2} e^{-n h}\right)$ is plotted with a yellow line.
The same result is obtained by taking into account the first two orders of approximation in (\ref{eq:diffchi}),
$ \Delta = \Delta_1 + \Delta_2, \quad \Delta_1 = - 2, \quad \Delta_2 = e^{-n h} $. 
For $N = 2$ the difference does not grow exponentially and $h' = h - h = 0$.

\vspace{1em}
\noindent {\it Dimension $N = 3$}
\vspace{1em}

\begin{figure}
\begin{center}
\includegraphics[width=0.7\linewidth]{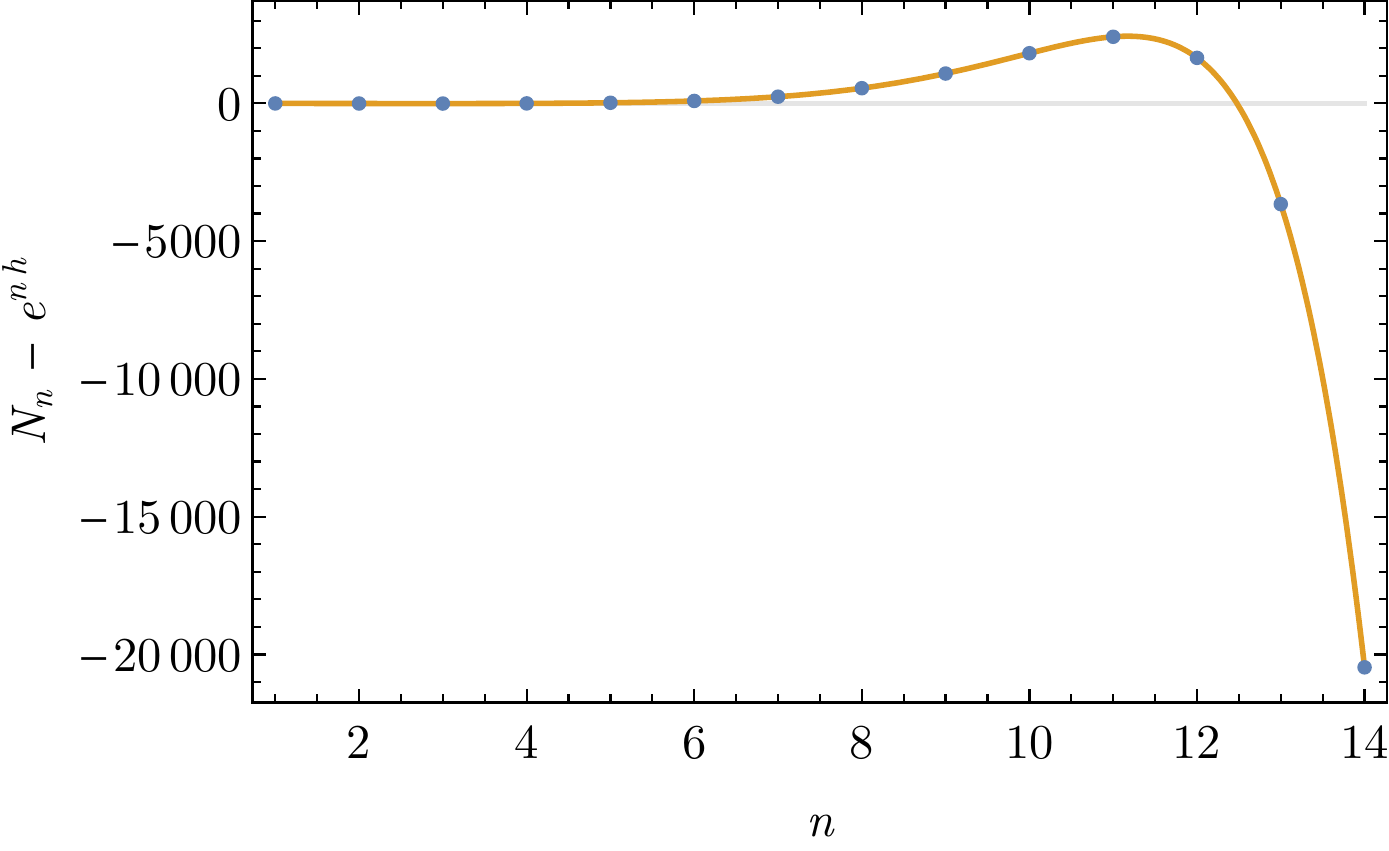}	
\caption{Difference between $\cN_n$ and $e^{n h}$ for $N = 3$.
It oscillates with $n$ as a result of complex eigenvalues of matrix $T$.
The amplitude grows as $e^{\frac{n h}{2}}$. }
\label{fig:diff3}	
\end{center}
\end{figure}

\begin{figure}
\begin{center}
\includegraphics[width=0.7\linewidth]{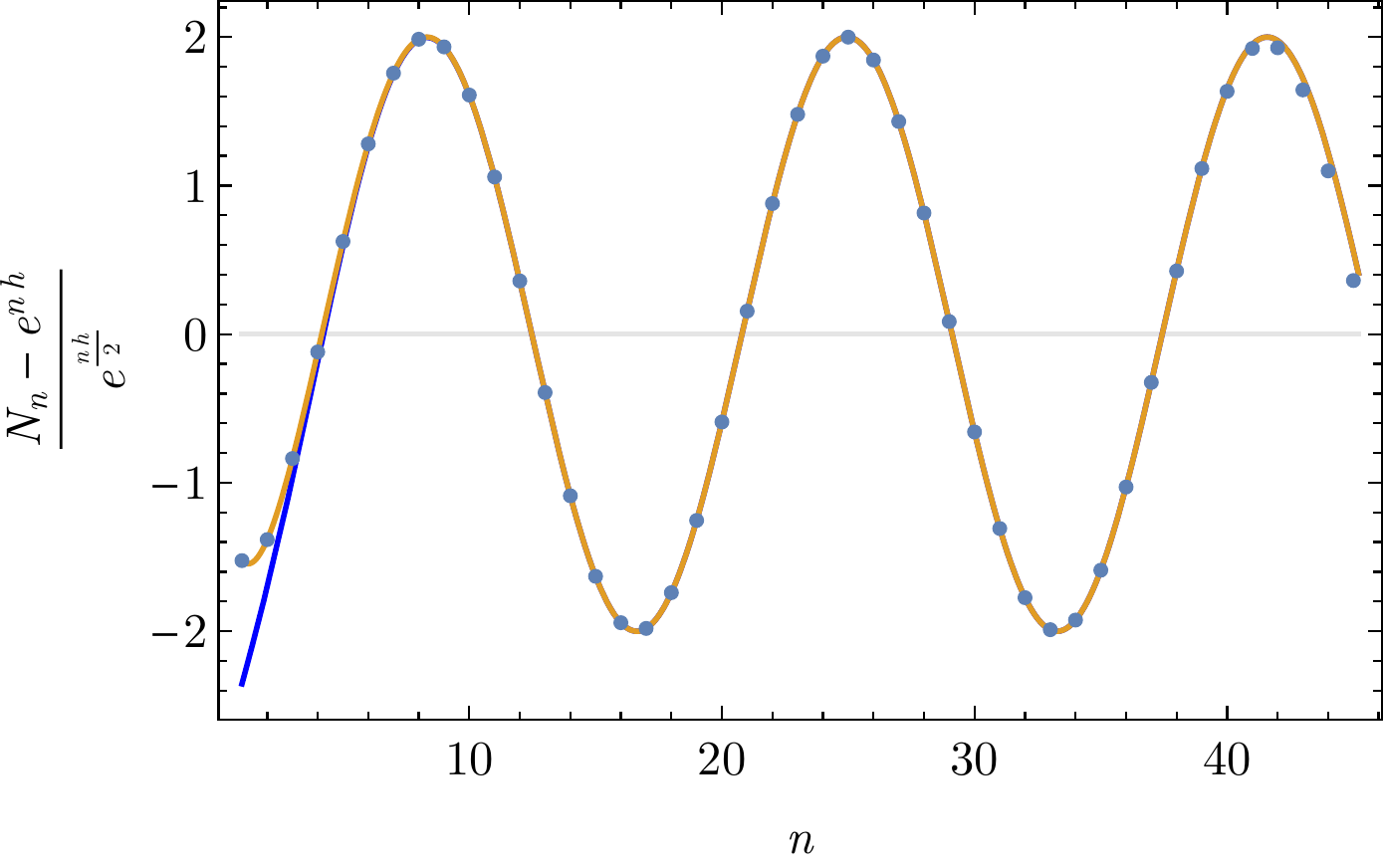}	
\caption{The factor $c(n)$ for $N = 3$.
Blue dots were obtained from calculation of the number of trajectories.
Yellow line plots the analytical solution.
The oscillatory behavior is a result of two eigenvalues which are complex conjugate.
The blue line is the first order approximation $\Delta_1 e^{- n h'}$}
\label{fig:diff3B}	
\end{center}
\end{figure}

In three dimensions the behavior of $\Delta$ is more interesting, 
but still can be calculated analytically.
In this case, the matrix $T$ has one real eigenvalue $\lambda_1 = e^h > 1$ and a complex pair of conjugate eigenvalues
$\lambda_2 = e^{-\frac{h}{2} + i \varphi}, \lambda_3 = \bar{\lambda}_2 = e^{-\frac{h}{2} - i \varphi}$.
Their product fulfills a condition $\lambda_1 \lambda_2 \lambda_3 = \Det T = 1$.
The above values allow direct calculation of the number of points $\cN_n$ and its deviation from $e^{n h}$,
\begin{equation}\label{eq:Nndiff3}
	\cN_n - e^{n h} = - 2 e^{\frac{n h}{2}} \left(1 - e^{-n h}\right) \cos(n \varphi) - e^{-n h}.
\end{equation}
The dominating exponential behavior $e^{\frac{n h}{2}}$ is in agreement with (\ref{eq:hprim}), $h' = h - \frac{h}{2} = \frac{h}{2}$.
For $N = 3$ the exponents are $h \approx 1.406, h' \approx 0.703$.

The difference (\ref{eq:Nndiff3}) is shown in Fig. \ref{fig:diff3}, the exponential growth $e^{n h'}$ of the amplitude and 
beginning of oscillatory behavior due to $\cos (n \varphi)$ are visible.
The blue dots were obtained by counting the number of points on periodic trajectories
and yellow line is the functional form (\ref{eq:Nndiff3}). 
The pair of complex conjugate eigenvalues generate the $\cos (n \varphi)$ term, which is real.
Its oscillatory character is clearly visible in the plot of 
factor $c(n) = \frac{\cN_n - e^{n h}}{e^{n h'}} = - 2 \left(1 - e^{-n h}\right) \cos(n \varphi) - e^{-3 n h / 2}$ (yellow line) in Fig. \ref{fig:diff3B}.
Approximations given by (\ref{eq:diffchi}) are
\[ \Delta_1 = - 2 e^{\frac{n h}{2}} \cos n \varphi - 1,\quad \Delta_2 = 1 + 2 e^{-\frac{n h}{2}} \cos n \varphi,\quad \Delta_3 = - e^{-\frac{n h}{2}} . \]
The first order approximation $\frac{\Delta_1}{e^{n h'}}$ is drawn with a blue line in Fig. \ref{fig:diff3B}.
It becomes indistinguishable from the exact solution for $n > 4$. 

\vspace{1em}
\noindent {\it Matrices of Higher Dimension}
\vspace{1em}

Although, the difference $\cN_n - e^{n h}$ is a simple function of eigenvalues $T$,
the lack of knowledge about their values hinders derivation of an analytical form.
For the sake of convenience we sort $\chi_i$'s in a descending order with respect to their modulus,
\[ 1 > |\chi_1| \geq |\chi_2| \geq \cdots \geq |\chi_N| > 0. \]
Depending on the position of eigenvalues with respect to the unit circle,
various patterns of the factor $c(n)$ can be observed:

\begin{figure}
\begin{center}
\includegraphics[width=0.7\linewidth]{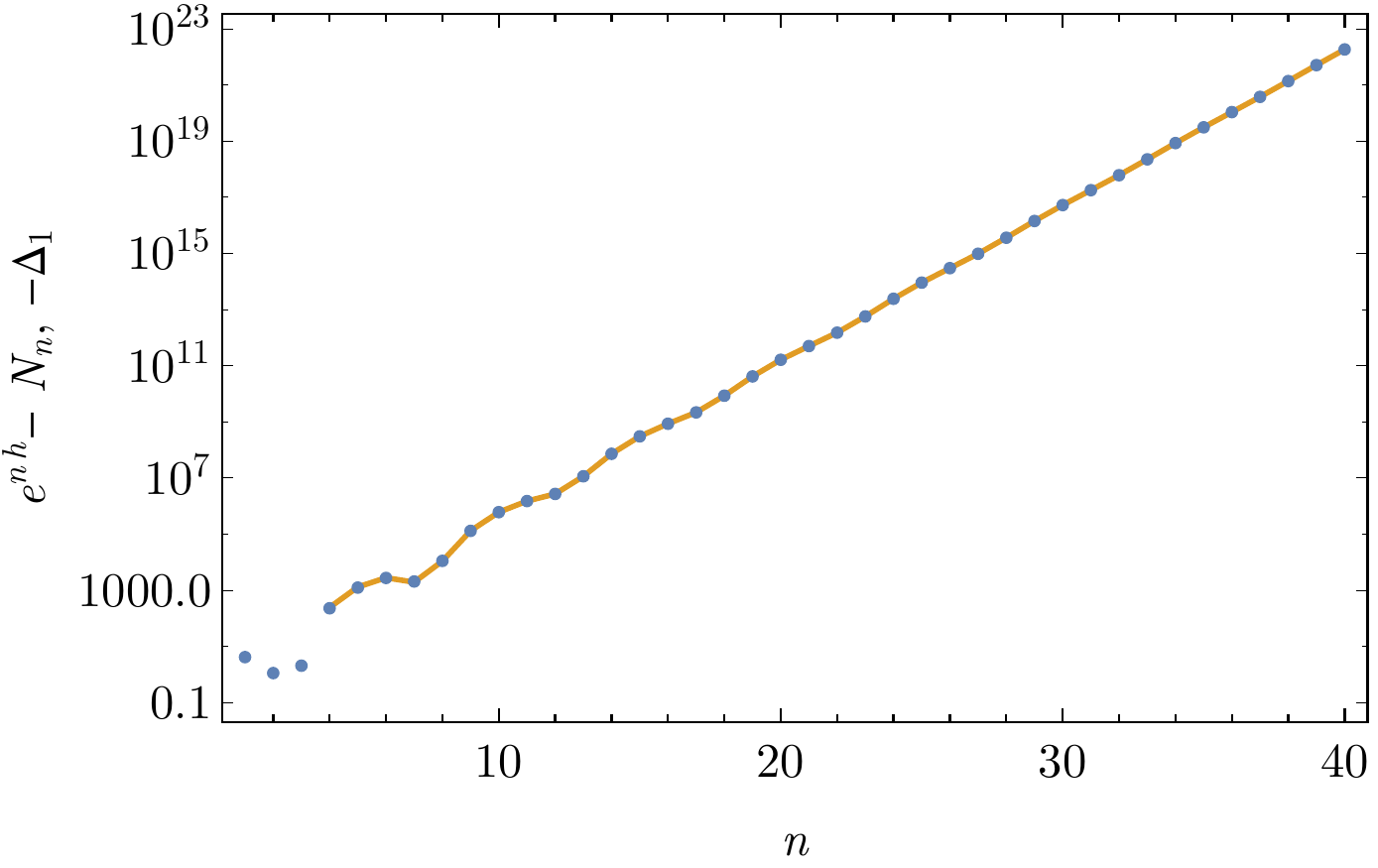}	
\caption{Difference between $e^{n h}$ and $\cN_n$ for $N = 4$ presented as blue dots on a logarithmic scale.
An exponential growth $e^{h' n}$ is well visible.
The approximation $\Delta_1$ is very accurate for $n > 4$ (yellow line).}
\label{fig:diff4}	
\end{center}
\end{figure}

\begin{figure}
\begin{center}
\includegraphics[width=0.7\linewidth]{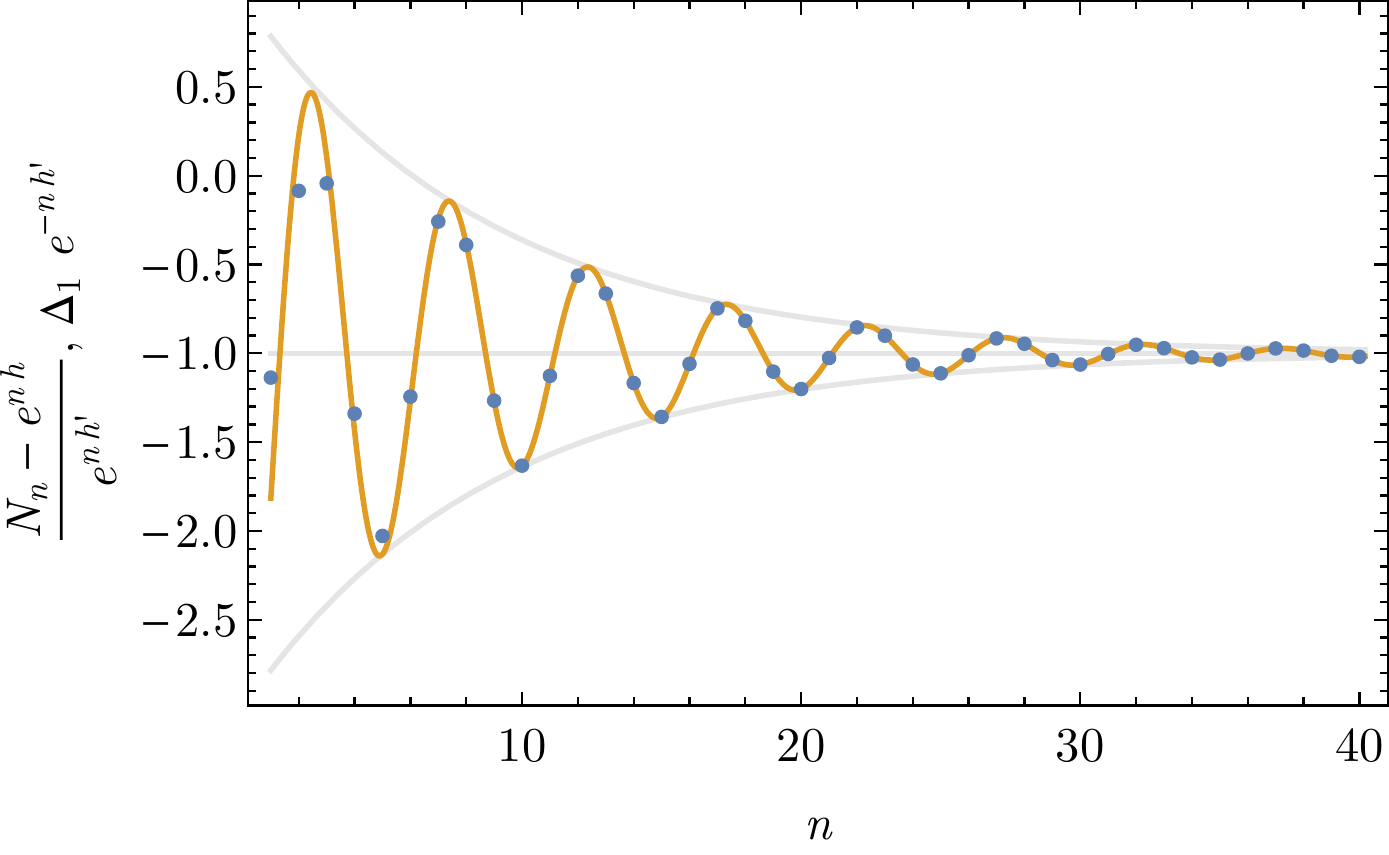}	
\caption{Damped oscillatory behavior of the scaled difference $\frac{\cN_n - e^{n h}}{e^{n h'}}$
(blue dots) for $N = 4$ is a result of a single real eigenvalue with modulus closest to $1$ (largest $\chi_i$) 
and a conjugate pair of complex eigenvalues ($\chi$) with lower modulus.
Scaled approximation $\Delta_1 e^{-n h}$ given by eq. (\ref{eq:diffchi}) is plotted with a yellow line.}
\label{fig:diff4b}	
\end{center}
\end{figure}

\noindent {\it  Damped oscillations.}
The difference $\Delta$ for $N = 4$ is plotted with blue dots in Fig. \ref{fig:diff4}.
As predicted by (\ref{eq:hprim}), it grows exponentially as $e^{n h'}, \ h' \approx 1.28$.
The entropy in this case is $h \approx 1.81$.
The factor $c(n)$ is shown in Fig. \ref{fig:diff4b}.
In both plots the yellow line denotes the first order approximation $\Delta_1$.
It gives very fast an accurate result.
The visible damped oscillatory behavior of $c(n)$ is a result of a single real value
$\chi_1$ with largest modulus (corresponding eigenvalue closest to unit circle) and a conjugate pair of complex eigenvalues $\chi_2 = \bar{\chi}_3$ with smaller modulus.
The amplitude of oscillations declines as $\frac{|\chi_2|^n}{|\chi_1|^n}$ which is drawn with a gray envelope in Fig. \ref{fig:diff4b}.
In this case as the oscillations fade out, the largest $\chi_1$ dominates and the factor $c(n)$ converges to a constant, 
\[ c(n) =  \frac{\cN_n - e^{n h}}{e^{n h'}} \underset{n \to \infty}{\longrightarrow} -1. \]
Similar behavior is observed also for $N = 8$.

\begin{figure}
\begin{center}
\includegraphics[width=0.7\linewidth]{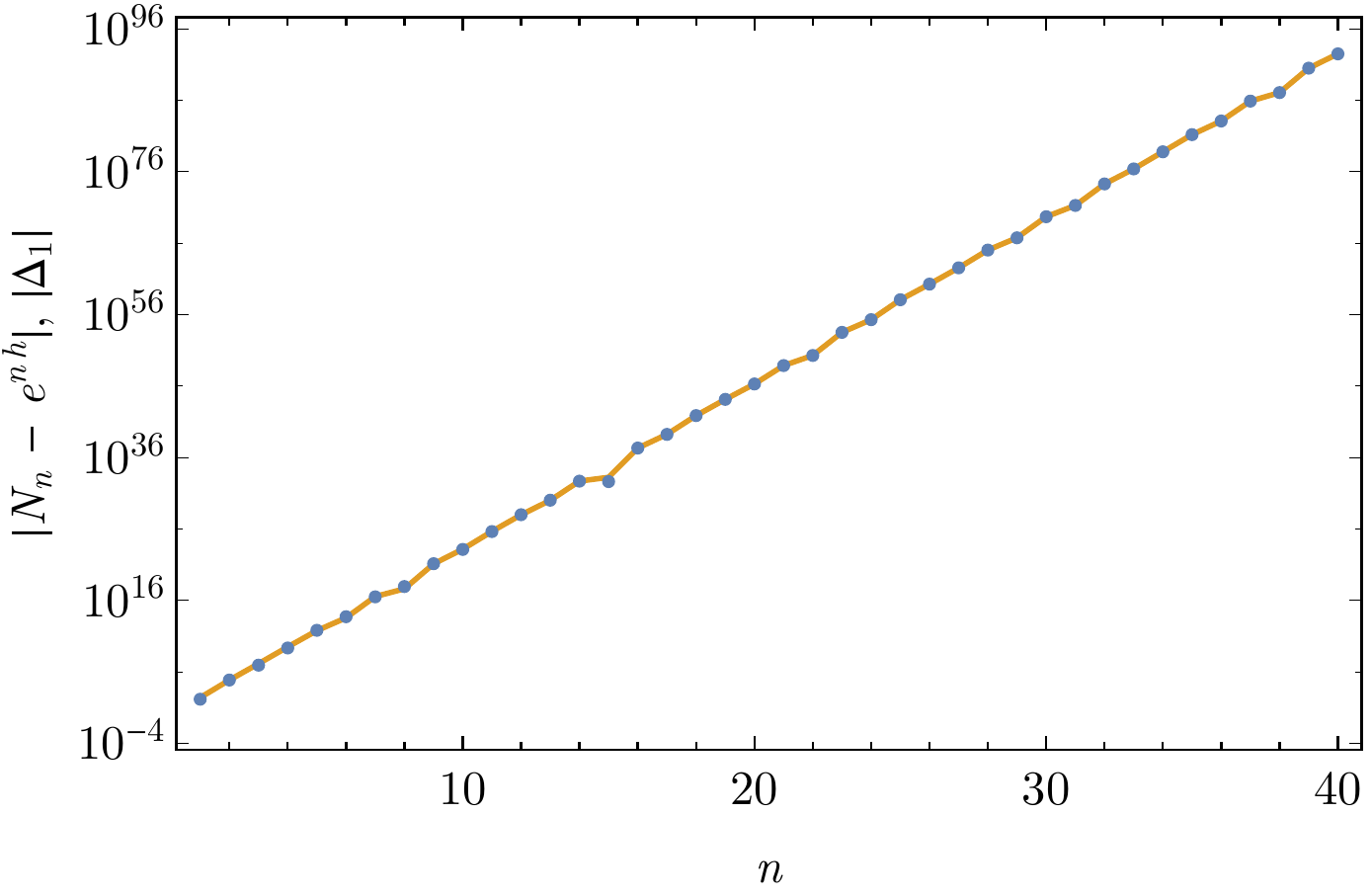}	
\caption{
Blue dots denote the difference between $\cN_n$ and $e^{n h}$ for $N = 10$.
It grows exponentially as $e^{h' n}$, $h' \approx 5.3$.
The approximation $\Delta_1$ is drawn with a yellow line.}
\label{fig:diff10}	
\end{center}
\end{figure}

\begin{figure}
\begin{center}
\includegraphics[width=0.7\linewidth]{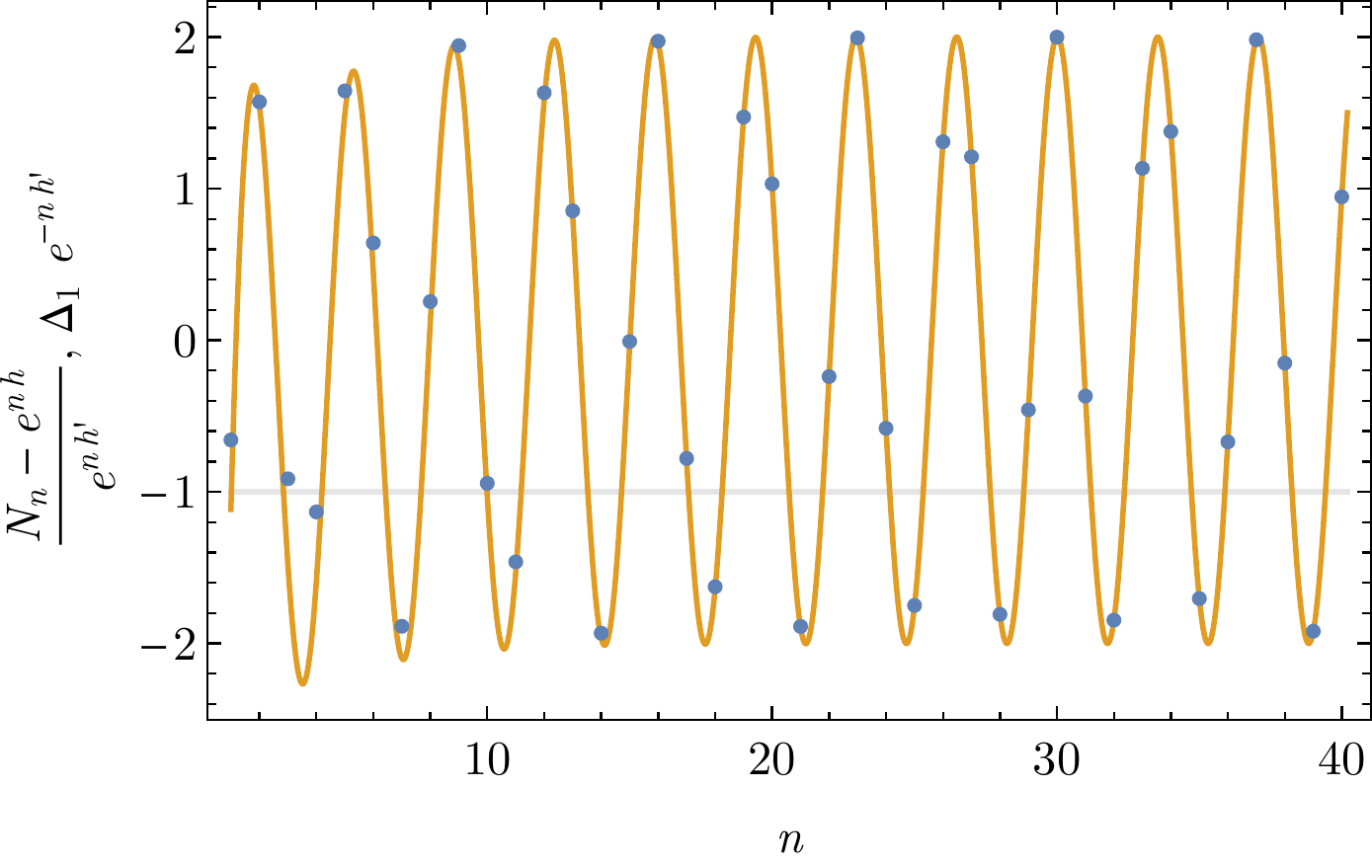}	
\caption{
Blue dots denote the factor $c(n)$ for $N = 10$.
The first order approximation $\Delta_1 e^{- n h'}$ is drawn with a yellow line.
The oscillations are produced by an isolated pair of complex conjugate values $\chi_1, \chi_2$ with largest modulus.}
\label{fig:diff10b}	
\end{center}
\end{figure}
\noindent {\it Oscillations.}
The damped oscillatory behavior of $c(n)$ present for $N = 4$ is more an exception than a general rule.
Fig. \ref{fig:diff10} presents the difference $\Delta$ for $N = 10$.
As in all cases it grows exponentially as $e^{n h'}, \ h' \approx 5.32$ while the entropy is $h \approx 5.57$.
The yellow line denotes the first order approximation $\Delta_1$ which again is remarkably accurate for $n > 4$.
Fig. \ref{fig:diff10b} shows the factor $c(n)$ for $N = 10$ with blue dots
and the approximation $\Delta_1 e^{- n h'}$ with a yellow line.

For $N = 10$ there is an isolated pair of complex conjugate values $ \chi_1 $ and $ \chi_2 = \bar{\chi_1} $ with the largest modulus.
A gap between the next element $|\chi_3|$ diminishes a contribution from the rest of eigenvalues. 
Such setting generates the oscillatory behavior of $c(n)$ with a roughly constant amplitude.
Similar behavior appears for most values of $N$.

\begin{figure}
\begin{center}
\includegraphics[width=0.7\linewidth]{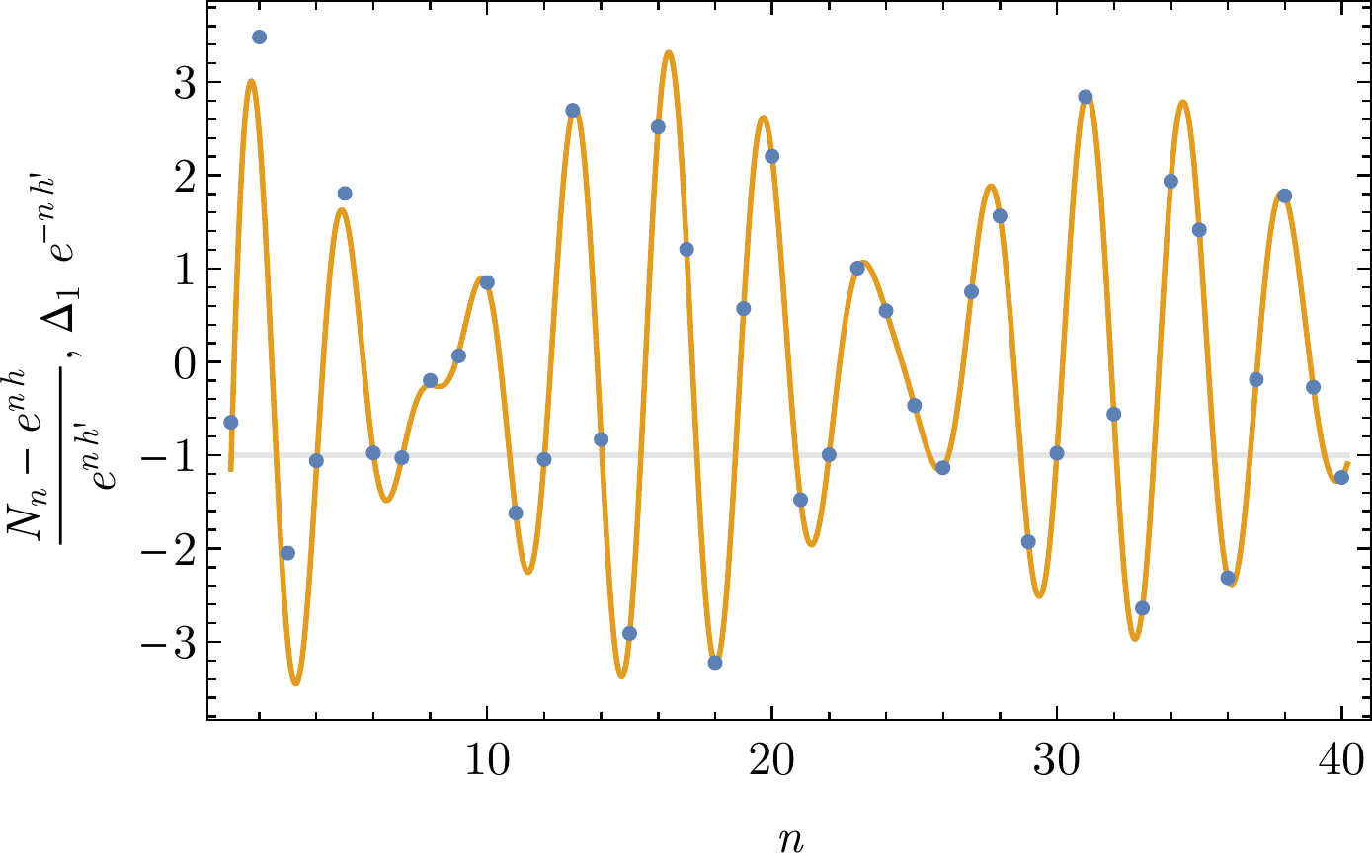}	
\caption{
Blue dots denote the factor $c(n)$ for $N = 14$.
The first order approximation $\Delta_1 e^{- n h'}$ is drawn with a yellow line.
Two pairs of complex conjugate $\chi_i$ with very largest but similar modulus
generates a beating.}
\label{fig:diff14b}	
\end{center}
\end{figure}

\noindent {\it Beating.}
\vspace{1ex}
It might happen that a gap between the two pairs of complex conjugate values 
with the largest modulus is very small.
The contribution from the second pair diminishes very slowly
and their interference generates a beating due to a difference in the argument.
Such situation is observed for $N = 14$ and roughly appears every sixth value of dimension $N$.
Fig. \ref{fig:diff14b} presents the beating of $c(n)$ with blue dots,
and the next to leading approximation with a yellow line.


\section{\it Acknowledgement }
The authors would like to thank Jan Ambjorn for stimulating discussions and support.  This work was supported in part by the European Union's Horizon 2020
research and innovation programme under the Marie Sk\'lodowska-Curie grant agreement No 644121. AG acknowledges support by the National Science Centre, Poland under grant no. 2015/17/D/ST2/03479.

\section*{\it Appendix }
\label{appx:Nndet}

The number of all points $\cN_n$ on the periodic trajectories of the period $n$ is given in (\ref{numbers}),
\[ \cN_n = \left| \Det \ T^n - 1 \right| = \prod_i \left| \lambda_i^n - 1 \right| = 
	\prod_{|\lambda_\alpha| < 1} \left| \lambda_\alpha^n - 1 \right| \prod_{|\lambda_\beta| > 1} \left| \lambda_\beta^n - 1 \right|.  \]
Because $T$ is a real matrix its complex eigenvalues will always occur in complex conjugate pairs. 
The large $n$ behavior is dominated by the largest eigenvalues (with respect to modulus) and is described by the entropy $h$,
\[ \cN_n \sim e^{n h} = \prod_{|\lambda_\beta| > 1} \left| \lambda_\beta \right|^n, \quad h = \sum_{|\lambda_\beta| > 1} \ln |\lambda_\beta|. \]
Before we derive an expression for the difference $\cN_n - e^{n h}$, let us first calculate a ratio
\[ \frac{\cN_n}{e^{n h}} = \prod_{|\lambda_\alpha| < 1} \left| \lambda_\alpha^n - 1 \right| \prod_{|\lambda_\beta| > 1} \frac{| \lambda_\beta^n - 1 |}{| \lambda_\beta^n|}
=  \prod_{|\lambda_\alpha| < 1} \left| \lambda_\alpha^{n} - 1 \right| \prod_{|\lambda_\beta| > 1} \left|1 - \lambda_\beta^{-n} \right|  \]
Because $|\lambda_\alpha| < 1$ and complex $\lambda_\alpha$ appear in complex conjugate pairs, below product is necessarily real and positive,
$ \prod_{|\lambda_\alpha| < 1} (1 - \lambda_\alpha^n) = \prod_{|\lambda_\alpha| < 1} |\lambda_\alpha^n - 1| . $
The same argument holds for $\lambda_\beta^{-1}$ which leads to
\[ \frac{\cN_n}{e^{n h}} =  \prod_{|\lambda_\alpha| < 1} (1 - \lambda_\alpha^n) \prod_{|\lambda_\beta| > 1} (1 - \lambda_\beta^{-n}) = \prod_{i} (1 - \chi_i^{n}), \]
where
\[ 
\chi_i = \left\{
	\begin{array}{rl}
		\lambda_i	& \text{if } |\lambda_i| < 1,\\
		\frac{1}{\lambda_i}	& \text{if } |\lambda_i| > 1.
	\end{array}		\right.
\]
The final expression is 
\[ \cN_n - e^{n h} = e^{n h}\left[ \prod_{i} (1 - \chi_i^{n}) - 1 \right] . \]

\vfill

\end{document}